\documentclass[12pt]{iopart}

\usepackage{iopams}  
\usepackage{graphicx}
\usepackage{bm}
\usepackage{color}
\usepackage{ulem}
\usepackage{array}
\usepackage{cite}

\begin{document}

\title{Spin-1/2 kagome-lattice antiferromagnets Cs$_2$Cu$_3$SnF$_{12}$ and Rb$_2$Cu$_3$SnF$_{12}$}

\author{Hidekazu Tanaka}

\address{Center for Entrepreneurship Education, Tokyo Institute of Technology, Midori-ku, Yokohama 226-8502, Japan}
\ead{tanaka.h.ag@m.titech.ac.jp}
\vspace{10pt}
\begin{indented}
\item[\today]
\end{indented}

\begin{abstract}
Spin-1/2 kagome-lattice Heisenberg antiferromagnet is theoretically known to have a quantum spin liquid (QSL) ground state, one of the frontiers of condensed matter physics. The search for the model substances has been continued for a long time, and many candidate substances have been reported. Most of them are hydroxide minerals. Here, I review the static and dynamic properties of non-mineral kagome-lattice antiferromagnets Cs$_2$Cu$_3$SnF$_{12}$ and Rb$_2$Cu$_3$SnF$_{12}$. X-ray and magnetic susceptibility measurements revealed that the kagome lattice in Cs$_2$Cu$_3$SnF$_{12}$ is approximately uniform, while that in Rb$_2$Cu$_3$SnF$_{12}$ is modified so that four different nearest-neighbor exchange interactions are arranged to form a pinwheel. Both systems are free of lattice disorder that gives rise to the exchange randomness. As large single crystals are obtainable, details of magnetic excitations can be studied by neutron scattering experiments. Cs$_2$Cu$_3$SnF$_{12}$ has a ${\bm q}\,{=}\,0$ ordered ground state due to the DM interaction. Magnetic excitations consist of four single magnon modes and an intense broad continuum, which conventional linear spin wave theory cannot explain. Based on a minimal model expressed by the Heisenberg term and the Dzyaloshinskii-Moriya (DM) term with the ${\bm D}$ vector perpendicular to the kagome layer, recent theories successfully describe the features of the magnetic excitations. From the experimental and theoretical results, it is deduced that the ground state of Cs$_2$Cu$_3$SnF$_{12}$ is given by the superposition of the ordered component and the QSL component.
The ground state of Rb$_2$Cu$_3$SnF$_{12}$ is a so-called pinwheel valence bond solid with an excitation gap. Magnetic excitations are composed of singlet-triplet modes and an intense high-energy continuum. From the analysis of the triplet modes, the magnitudes of the four kinds of exchange interaction and the DM interaction are determined. The high-energy excitation continuum is observed in common with Cs$_2$Cu$_3$SnF$_{12}$ and Rb$_2$Cu$_3$SnF$_{12}$, irrespective of their ground states, and thus, it is a common characteristic of spin-1/2 kagome-lattice antiferromagnets.
\end{abstract}

%
\vspace{2pc}

\section{Introduction}\label{Intro}
Quantum many-body effects in frustrated quantum magnets produce exotic ground states and excitations~\cite{Balents_2010,Broholm}. Spin-1/2 triangular-lattice Heisenberg antiferromagnet (TLHAF) with the nearest-neighbor (NN) exchange interaction is the simplest and prototypical frustrated quantum magnet. The classical ground state of the TLHAF is an ordered state with a 120$^{\circ}$ structure. Since Anderson~\cite{Anderson_1973} proposed a resonating-valence-bond (RVB) spin liquid state without a long-range ordering as the ground state of the spin-1/2 TLHAF, significant effort has been made to elucidate the nature of the ground state. The current theoretical consensus is that the ground state is the 120$^{\circ}$ ordered state with a significantly reduced sublattice magnetization~\cite{Miyashita_1984,Huse,Jolicoeur,Bernu,Singh,White,Gotze,Li2}. Although the ground state is ordered, the spin-1/2 TLHAF exhibits a magnetization plateau at one-third of the saturation magnetization induced by quantum fluctuation\cite{Nishimori,Chubokov,Nikuni,Honecker,Alicea,Farnell,Sakai,Hotta,Yamamoto1,Starykh_Review,Sellmann,Coletta}. This macroscopic quantum many-body effect was experimentally observed in many spin-1/2 Heisenberg-like triangular-lattice antiferromagnets (TLAFs), although its emergence is modified by spatial anisotropy, exchange anisotropy, and interlayer coupling~\cite{Ono3,Ono4,Fortune,Shirata,Susuki,Okada,Sera,Kojima,Xing,Ranjith,Ding,Yamamoto4}.

Theoretically, magnetic excitations in spin-1/2 TLHAF have been studied using various approaches~\cite{Ghioldi,Ghioldi2,Ghioldi3,Chi,Syromyatnikov2,Ferrari,Zhang,Starykh,Zheng,Chernyshev,Mezio,Mourigal}. It was found that the dispersion relations are significantly different from those of the linear spin wave theory, e.g., the excitation energy is significantly renormalized downward in a large area of the Brillouin zone (BZ)~\cite{Starykh,Zheng,Chernyshev,Mezio,Mourigal}. The dispersion curve shows a roton-like local minimum at the M point~\cite{Zheng,Mezio,Verresen}. The understanding of magnetic excitations has progressed rapidly in the last decade owing to the inelastic neutron scattering (INS) experiments on  Ba$_3$CoSb$_2$O$_9$~\cite{Ma,Ito,Kamiya,Macdougal,Zhou}. The experiments verified the above theoretical predictions and revealed that the excitation spectrum has a structured continuum extending to the high-energy region, the intensity of which is so strong that it cannot be explained in terms of two-magnon excitations. The experimental results demonstrated that fractional excitations such as spin-1/2 excitations spinons dominate the spectrum. These experimental results encouraged theoretical studies. Although there is still disagreement between theory and experimental results, fermionic or bosonic spinon theory~\cite{Ghioldi,Ghioldi2,Ghioldi3,Ferrari,Zhang}, tensor-network~\cite{Chi}, and bond operator approaches~\cite{Syromyatnikov2} captured well the characteristics of the excitation spectrum observed in Ba$_3$CoSb$_2$O$_9$. These theoretical and experimental results of magnetic excitations strongly suggest that the ground state has a sizable quantum spin liquid (QSL) component because spinons are characteristic excitations of the QSL. The studies of magnetic excitations are essential to understanding the nature of the ground state in frustrated quantum magnets.

Spin-1/2 kagome-lattice Heisenberg antiferromagnet (KLHAF) with the NN exchange interaction is another prototypical frustrated quantum magnet. Extensive theoretical studies demonstrated that the ground state is quantum-disordered~\cite{Zeng1,Sachdev,Chalker,Elstner,Nakamura,Zeng2,Lecheminant,Waldtmann,Mila,Mambrini,Syromyatnikov,Jiang}. Although valence bond solids (VBSs) described by a periodic array of singlet dimers were also discussed previously\cite{Nikolic,Budnik,Singh1,Singh2,Yang,Hwang1}, most recent theories support the QSL ground state, including the gapless $U(1)$ Dirac spin liquid state~\cite{Hastings,Ran,Hermele,Iqbal,Liao,He,Zhu,Jiang2} and gapped $Z_2$ spin liquid state~\cite{Jiang3,Yan,Depenbrock,Jiang4,Mei}. The ground state of the spin-1/2 KLHAF qualitatively differs from the ordered ground state for the large spin KLHAF, where quantum fluctuation stabilizes the $\sqrt{3}\,{\times}\,\sqrt{3}$ structure~\cite{Chubukov2,Reimers}.
Theoretical studies of magnetic excitations in spin-1/2 KLHAF appear less detailed, as compared to the case of spin-1/2 TLHAF, because the ground state nature is unclear.
An intriguing theoretical prediction is the emergence of successive magnetization plateaus~\cite{Hida,Honecker_kagome,Nishimoto,Capponi,Schnack,Plat}. However, unlike the 1/3 plateau in the spin-1/2 TLHAF, the plateaus in the spin-1/2 KAHAF melt rapidly by the slight temperature increase due to many excited states with different total spins energetically close to the plateau state~\cite{Morita_kagome}. Experimental studies of field-induced phase transitions, including magnetization plateaus, were reviewed in Ref.~\cite{Yoshida}. 

The predicted spin liquid ground state stimulated the search for spin-1/2 kagome-lattice antiferromagnet (KLAF), and many candidate compounds were reported~\cite{Mueller,Hiroi,Shores,Okamoto,Fak,Ishikawa,Fujihala,Sun,Sun2,Yoshida2,Puphal,Feng,Wang}. Most of them are minerals of hydroxides. Among them, Herbertsmithite ZnCu$_3$(OH)$_6$Cl$_2$ has been studied in the most detail~\cite{Mendels,Helton,Bert,Lee,Imai,Olariu,Vries,Mendels2,Han2,Fu,Zorko,Khuntia}. Whether its ground state is gapless or gapped has been controversially discussed~\cite{Mendels,Imai,Mendels2,Fu}. The magnetic excitations in Herbertsmithite were investigated via INS experiments using single crystals~\cite{Han,Han3}. The excitation spectrum is broad and appears almost featureless. The experimental results stimulated theoretical studies on magnetic excitations~\cite{Punk,Zhang2,Zhu,Ferrari2,Prelovsek}. 
However, synthesized Herbertsmithite has a problem in that 15\% of Zn$^{2+}$ sites are substituted for Cu$^{2+}$ ions~\cite{Han2,Han3}. A similar substitution for Cu$^{2+}$ ions has also been reported in Cu$_3$Zn(OH)$_6$FBr~\cite{Wang}. Due to the Jahn-Teller effect, the Cu(OH)$_6$ octahedron in the Zn$^{2+}$ layer is distorted to break the local trigonal symmetry of the octahedron. This gives rise to the disorder of OH$^-$ ions that mediate superexchange interaction between Cu$^{2+}$ ions in the kagome layer and leads to the exchange randomness in the kagome layer. Hence, the magnetic properties of the kagome layer are not independent of the Cu$^{2+}$ substitution for Zn$^{2+}$ ions. Recent theories demonstrated that exchange randomness produces a disordered ground state called a random singlet state in frustrated quantum magnets and creates indistinct magnetization plateaus and excitation spectrums~\cite{Watanabe,Kawamura2014,Shimokawa,Kawamura2019}. The synthesis of non-copper-doped Herbertsmithite is desired.

The information on the magnetic excitations is essential to know the magnetic interactions and the ground state nature of spin-1/2 KLHAFs. The INS experiments on Vesignieite~\cite{Boldrin} and Kapellasite~\cite{Fak} have been conducted using powder samples. These experiments revealed that the dominant exchange interactions are the third neighbor or the diagonal interaction, not the NN interaction. However, it is difficult to observe directly the detailed structure of the excitation spectrum from powder scattering data.

This article reviews the static and dynamic properties of spin-1/2 KLAFs Cs$_2$Cu$_3$SnF$_{12}$ and Rb$_2$Cu$_3$SnF$_{12}$ and related theories. Although kagome lattices are distorted, their advantages are (1) no exchange randomness due to the lattice or charge disorder, (2) the NN exchange is antiferromagnetic and dominant so that the system can be described as a spin-1/2 KLAF with the NN exchange interaction, (3) large single crystals can be obtained, and (4) no elements inappropriate for neutron scattering are contained. 

For Cs$_2$Cu$_3$SnF$_{12}$, the lattice distortion is slight, and the system can approximate a uniform KLAF. The exchange interaction was evaluated as $J/k_{\rm B}\,{=}\,240$\,K (${=}\,20.7$ meV)~\cite{Ono}. Cs$_2$Cu$_3$SnF$_{12}$ undergoes a magnetic phase transition at $T_{\rm N}\,{=}\,20.0$ K to a $\bm{q}\,{=}\,0$ state with positive chirality owing to the Dzyaloshinskii-Moriya (DM) interaction and the weak interlayer exchange interaction~\cite{Ono,Ono2,Matan3}. The magnetic excitation spectrum consists of single magnon branches and a broad continuum ranging from $0.15J$ to at least $2.5J$~\cite{Saito,Matan4}. The characteristics of the continuum are consistent with the results of fermionic spinon theories~\cite{Zhang2,Ferrari3}. The dispersion relations of the single magnon excitations were drastically changed from the result of linear spin theory~\cite{Kogure}. The results of magnetic excitations strongly suggest that a sizable spin liquid component remains in the ground state as the quantum fluctuation, similar to the spin-1/2 TLAF case, even though the ground state is ordered.

The chemical unit cell at room temperature for Rb$_2$Cu$_3$SnF$_{12}$ is $2a\times2a$ enlarged in the $ab$ plane; thus, there are four kinds of the NN exchange interaction $J_\alpha$ ($\alpha\,{=}\,1$-$4$) that form a pinwheel~\cite{Morita}. Owing to the exchange network, the magnetic ground state of Rb$_2$Cu$_3$SnF$_{12}$ is a pinwheel VBS with an excitation gap of $\Delta\,{=}\,2.4$ meV~\cite{Morita,Ono,Matan,Matan2}. From the analysis of the dispersion curves for singlet-triplet excitations, the exchange constants and the out-of-plane component of DM interaction were determined to be $J_1\,{=}\,18.6$\,meV, $J_2\,{=}\,0.95J_1$, $J_3\,{=}\,0.85J_1$, $J_4\,{=}\,0.55J_1$, and $D_\alpha^\parallel\,{=}\,0.18J_\alpha$~\cite{Matan,Matan2}. It was also found that a broad excitation continuum extends to a high energy of 2.5 times the average of the NN exchange constants, similar to the Cs$_2$Cu$_3$SnF$_{12}$ case~\cite{Saito}. The characteristics of the high-energy excitation continuum are common to both Cs$_2$Cu$_3$SnF$_{12}$ and Rb$_2$Cu$_3$SnF$_{12}$, irrespective of their ground states.

This review is organized as follows. Section~\ref{structure} describes the crystal structures and exchange networks in Cs$_2$Cu$_3$SnF$_{12}$ and Rb$_2$Cu$_3$SnF$_{12}$. In Section~\ref{sample}, details for sample preparation are provided. Section~\ref{static} shows the static magnetic properties investigated by magnetic susceptibility and specific heat measurements. Section~\ref{excitations} describes the magnetic excitations and related theories. The last Section is devoted to a conclusion.

\begin{figure}[t]
	\centering
	\includegraphics[width=13cm]{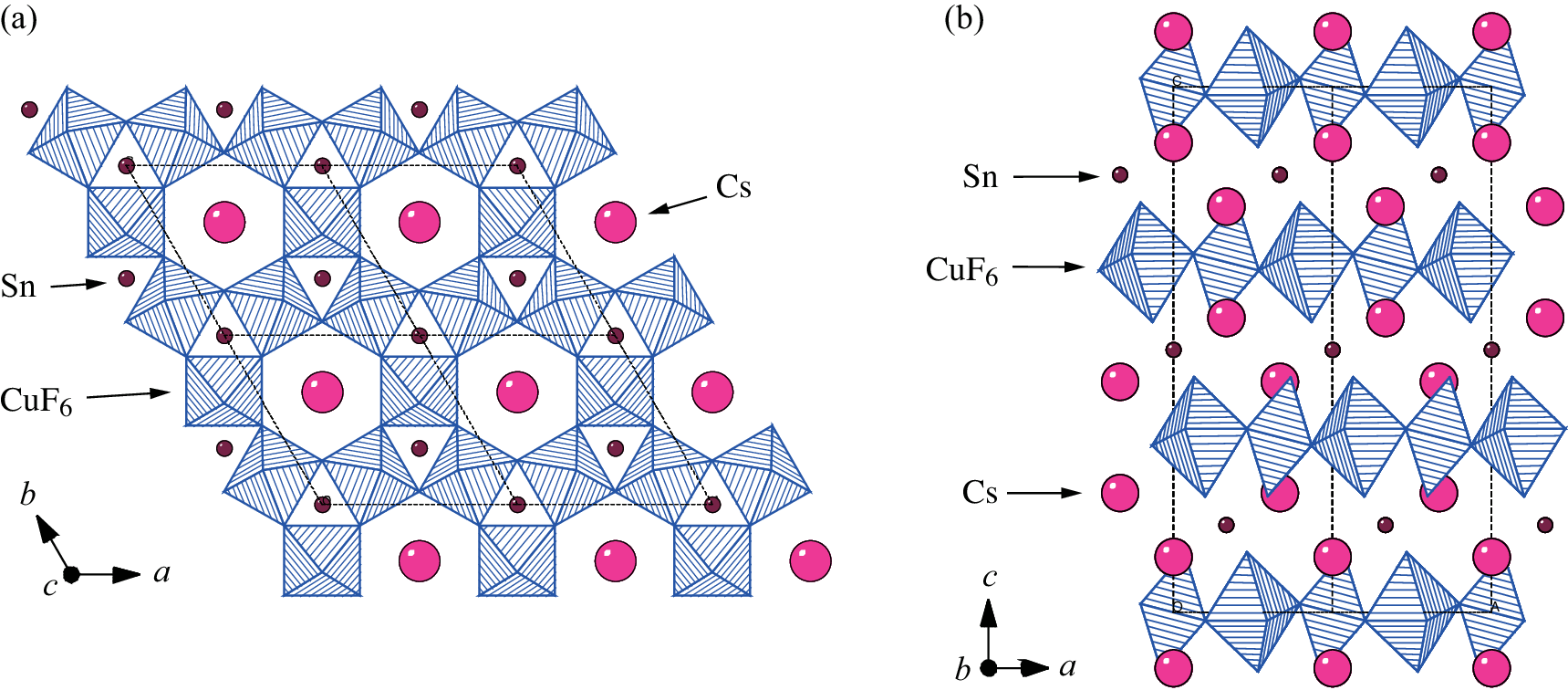}
	\caption{Room temperature crystal structures of Cs$_2$Cu$_{3}$SnF$_{12}$ viewed (a) along the $c$ axis and (b) along the $b$ axis. Dotted lines denote the chemical unit cells. Shaded blue octahedra are CuF$_6$ octahedra centered by Cu$^{2+}$ ions.}
	\label{fig:Cs_structure}
\end{figure} 

\begin{figure}[h]
	\centering
	\includegraphics[width=8cm]{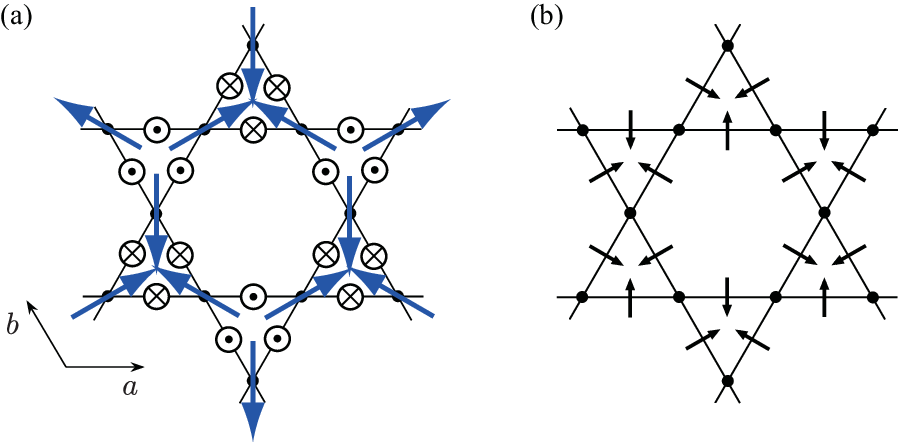}
	\caption{Arrangement of the $\bm{D}$ vectors for the DM interaction at room temperature in Cs$_2$Cu$_3$SnF$_{12}$. (a) The $c$ axis component $D^{\parallel}$ and (b) the $ab$ plane component $D^{\perp}$. Large blue arrows in (a) denote the $\bm{q}\,{=}\,0$ state with positive chirality stabilized by the DM interaction. Spin numbers $i$ and $j$ are given along the ${\bm a}$, ${\bm b}$, and ${-}\,{\bm a}\,{-}\,{\bm b}$ directions. Reprinted from Ref.~\cite{Ono} and modified.}
	\label{fig:DM}
\end{figure}  

\section{Crystal structures and exchange interactions}\label{structure} 

Cs$_2$Cu$_3$SnF$_{12}$ crystallizes in a trigonal structure (space group $R\bar{3}m$) at room temperature~\cite{Ono}, which is the same as the room-temperature structure of  Cs$_2$Cu$_3$ZrF$_{12}$ and Cs$_2$Cu$_3$HfF$_{12}$ \cite{Mueller}. 
Figure \ref{fig:Cs_structure} shows the room-temperature structure of Cs$_2$Cu$_3$SnF$_{12}$. CuF$_6$ octahedra centered by Cu$^{2+}$ ions are linked with sharing corners in the $ab$ plane. Magnetic Cu$^{2+}$ ions with spin-1/2 form a uniform kagome layer in the $ab$ plane. CuF$_6$ octahedra are elongated along the principal axes that make an angle of approximately 11$^{\circ}$ with the $c$ axis. Because the hole orbitals $d(x^2\,{-}\,y^2)$ of Cu$^{2+}$ ions lie in the kagome layer, the nearest neighbor superexchange interaction $J$ through F$^-$ ion in the kagome layer is antiferromagnetic and strong as $J/k_{\rm B}\,{=}\,240$\,K in the definition of ${\cal H}_{\rm ex}\,{=}\,J{\bm S}_i\,{\cdot}\,{\bm S}_j$~\cite{Ono}. The interlayer exchange interaction should be much smaller than $J$ because kagome layers of Cu$^{2+}$ are sufficiently separated by nonmagnetic Cs$^+$, Sn$^{4+}$, and F$^-$ layers. Hence, we can expect good two-dimensionality in Cs$_2$Cu$_3$SnF$_{12}$. 

Because there is no inversion center at the middle point of two neighboring magnetic ions in the kagome lattice, the Dzyaloshinsky-Moriya (DM) interaction expressed by ${\cal H}_{\rm DM}\,{=}\,{\bm D}_{ij}\,{\cdot}\,[{\bm S}_i\,{\times}\,{\bm S}_j]$ is allowed. Figure~\ref{fig:DM} shows the configuration of the ${\bm D}_{ij}$ vectors in Cs$_2$Cu$_3$SnF$_{12}$ derived from the symmetry of the crystal lattice~\cite{Ono}. The configuration of the ${\bm D}_{ij}$ vectors is the same as those in Fe- and Cr-based Jarosites discussed by Elhajal {\it et al.}\,\cite{Elhajal}. The $c$ axis (out-of-plane) component $D^{\parallel}$ stabilizes the ${\bm q}\,{=}\,0$ state with a positive chirality, while the $ab$ plane (in-plane) component $D^{\perp}$ acts to lift spins from the $ab$ plane. Neutron scattering experiments demonstrated that the ground state in Cs$_2$Cu$_3$SnF$_{12}$ is the ${\bm q}\,{=}\,0$ state with a positive chirality and that $D^{\parallel}$ is much larger than $D^{\parallel}$~\cite{Ono2}. 

\begin{figure}[h]
	\centering
	\includegraphics[width=10cm]{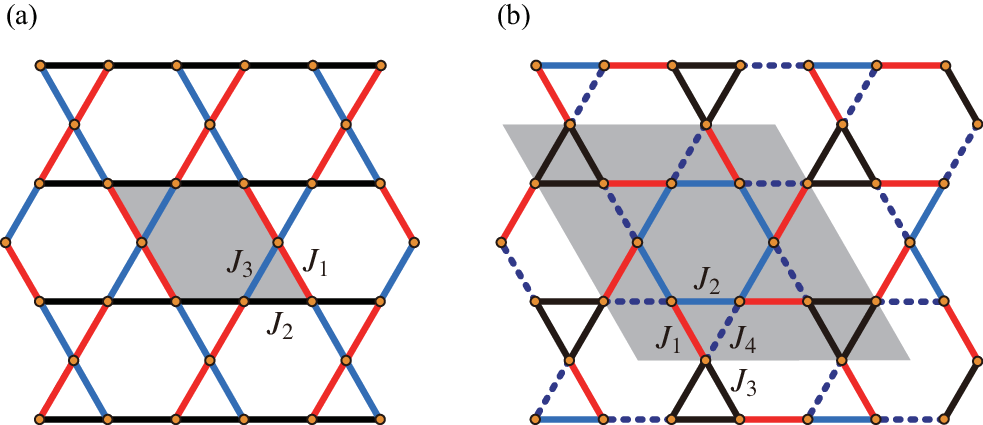}
	\caption{(a) Exchange network in the kagome layer of Cs$_2$Cu$_3$SnF$_{12}$ below the structural phase transition temperature $T_{\mathrm{t}}\,{=}\,185$ K~\cite{Matan3}. There are three kinds of NN exchange interactions $J_1, J_2$, and $J_3$. The magnitudes of $J_2$ and $J_3$ are estimated to be $J_2\,{\approx}\,0.95J_1$ and $J_3\,{\approx}\,0.90J_1$ from their bond angles Cu--F--Cu. (b) Exchange network at room temperature in the kagome layer of Rb$_2$Cu$_{3}$SnF$_{12}$. Pinwheels composed of exchange interactions $J_1$, $J_2$, and $J_4$ are connected by triangles of $J_3$ interactions~\cite{Morita,Matan}. Shaded areas in (a) and (b) are the unit cells for Cs$_2$Cu$_3$SnF$_{12}$ and Rb$_2$Cu$_3$SnF$_{12}$, respectively, at room temperature. Reprinted from Ref.~\cite{Saito}}
	\label{fig:ex_network}
\end{figure} 

\begin{figure}[t]
	\centering
	\includegraphics[width=8cm]{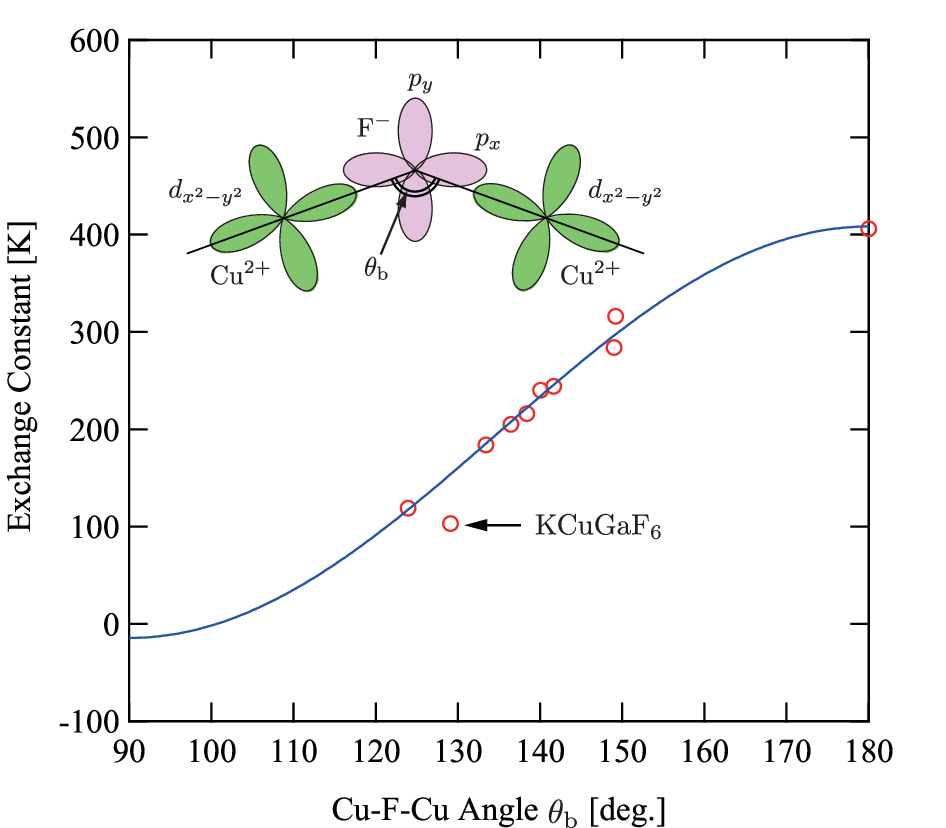}
	\caption{Superexchange interaction as a function of the Cu--F--Cu bond angle ${\theta}_{\rm b}$ in some copper fluoride magnets. The solid line denotes the fit by Eq.\,(\ref{eq:bond_angle}) with $J_{\rm AF}/k_{\rm B}\,{=}\,408.4$\,K and $J_{\rm F}/k_{\rm B}\,{=}\,{-}\,14.4$\,K. The data for KCuGaF6 is excluded from the fit because the Cu$^{2+}$ ion deviates significantly from the center of the octahedron. Inset shows the schematic orbital configuration for the superexchange path Cu--F--Cu and the bond angle ${\theta}_{\rm b}$.}
	\label{fig:exchange}
\end{figure}

Cs$_2$Cu$_3$SnF$_{12}$ undergoes a structural phase transition at $T_{\rm t}\,{=}\,185$ K~\cite{Ono}. Below $T_{\rm t}$, the crystal lattice is monoclinic ($P2_1/n$)~\cite{Downie2,Matan3}, and thus there are three kinds of nearest-neighbor exchange interactions $J_1, J_2$, and $J_3$ below $T_{\mathrm{t}}$, as shown in Fig.~\ref{fig:ex_network}~\cite{Downie2,Matan3}. It is known that the sign and magnitude of superexchange interaction are closely related to the Cu--F--Cu bond angle ${\theta}_{\rm b}$ (see the inset of Fig.~\ref{fig:exchange}). Table~\ref{tab:exchange_bond} shows the Cu--F--Cu bond angles ${\theta}_{\rm b}$ and exchange constants $J$ in some copper fluoride magnets. These data are summarized in Fig.~\ref{fig:exchange}. Since the superexchange interaction $J$ is antiferromagnetic and strong when ${\theta}_{\rm b}\,{=}\,180^{\circ}$ and ferromagnetic and weak when ${\theta}_{\rm b}\,{=}\,90^{\circ}$~\cite{Goodenough,Kanamori}, the exchange constant will be expressed as  
\begin{eqnarray}
	J=J_{\rm AF}\cos^2{\theta}_{\rm b}+J_{\rm F}\sin^2{\theta}_{\rm b},
	\label{eq:bond_angle}
\end{eqnarray}
where $J_{\rm AF}$ and $J_{\rm F}$ are exchange constants for ${\theta}_{\rm b}\,{=}\,180^{\circ}$ and $90^{\circ}$, respectively. The solid line denotes a fit with $J_{\rm AF}/k_{\rm B}\,{=}\,408.4$\,K and $J_{\rm F}/k_{\rm B}\,{=}\,{-}\,14.4$\,K. The small negative $J_{\rm F}$ shows that the exchange interaction is ferromagnetic and weak when ${\theta}_{\rm b}\,{=}\,90^{\circ}$, consistent with the Goodenough-Kanamori rule~\cite{Goodenough,Kanamori}. Equation\,(\ref{eq:bond_angle}) describes well the observed ${\theta}_{\rm b}$ dependence of $J$. Neutron diffraction experiment below $T_{\rm t}$ determined the bond angles of $J_1, J_2$, and $J_3$ to be $139.48^{\circ}$, $137.93^{\circ}$, and $136.31^{\circ}$, respectively~\cite{Matan3}. Using the result of the fit by Eq.\,(\ref{eq:bond_angle}), we can evaluate the magnitudes of these three exchange interactions as $J_2/J_1\,{\approx}\,0.95$ and $J_3/J_1\,{\approx}\,0.90$. This indicates that the exchange network below $T_{\rm t}$ in Cs$_2$Cu$_3$SnF$_{12}$ is approximately uniform. We also see that $J_1$ is almost the same as $J$ above $T_{\rm t}$ because the difference between bond angles of these couplings is only $0.6^{\circ}$ and the change in the magnetic susceptibility at $T_{\rm t}$ is little, as shown in Section 4~\cite{Ono}.

\begin{table}[h]
	\centering
	\caption{Cu--F--Cu bond angles ${\theta}_{\rm b}$ and exchange constants $J$ in some copper fluoride magnets.\vspace{2mm}}
	\label{tab:exchange_bond}
	\begin{tabular}{rrrr}
		\hline\hline
		Substance \ \  & ${\theta}_{\rm b}$\,[deg.] \ \ & $J$\,[K] \ \ & References \\\hline
		KCuF$_3$ \ \  & 180 \ \  & 406 \ \ & \cite{Satija,Nagler} \\
		Cs$_2$Cu$_3$CeF$_{12}$ ($J_1$) \ \  & 149.2 \ \  & 316 \ \ & \cite{Amemiya1} \\
		Cs$_2$Cu$_3$CeF$_{12}$ ($J_2$) \ \  & 149 \ \ & 297 \ \ & \cite{Amemiya1} \\
		Cs$_2$Cu$_3$ZrF$_{12}$ (${>}\,T_{\rm t}$) \ \  & 141.64 \ \ & 244 \ \ & \cite{Mueller,Ono} \\
		Cs$_2$Cu$_3$SnF$_{12}$ (${>}\,T_{\rm t}$) \ \ & 140.08 \ \ & 240 \ \ & \cite{Ono} \\
		Rb$_2$Cu$_3$SnF$_{12}$ ($J_1$) \ \ & 138.39 \ \ & 216 \ \ & \cite{Morita,Matan} \\
		Rb$_2$Cu$_3$SnF$_{12}$ ($J_2$) \ \ & 136.47 \ \ & 205 \ \ & \cite{Morita,Matan} \\
		Rb$_2$Cu$_3$SnF$_{12}$ ($J_3$) \ \ & 133.4 \ \ & 184 \ \ & \cite{Morita,Matan} \\
		Rb$_2$Cu$_3$SnF$_{12}$ ($J_4$) \ \ & 123.94 \ \ & 119 \ \ & \cite{Morita,Matan} \\
		KCuGaF$_6$ \ \ & 129.1 \ \ & 103 \ \ & \cite{Umegaki1,Umegaki2} \\
		\hline\hline 
	\end{tabular}
\end{table}

\begin{figure}[t]
	\centering
	\includegraphics[width=13cm]{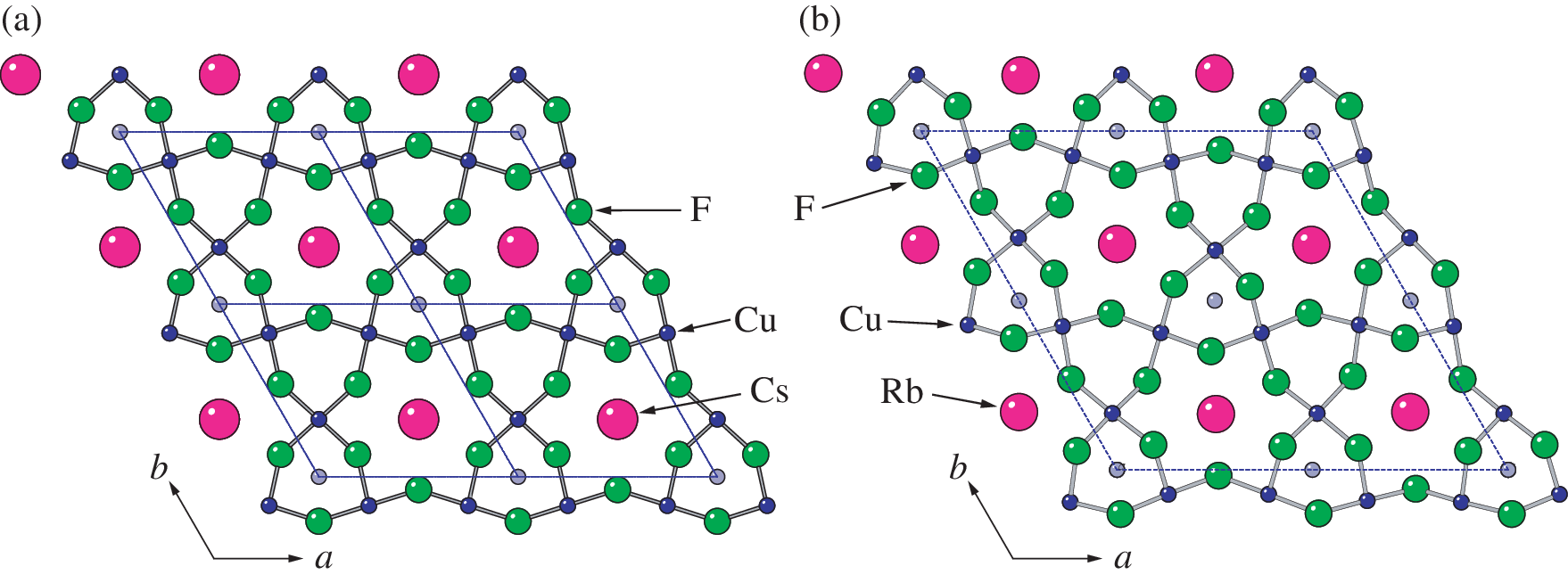}
	\caption{Room temperature crystal structures of (a) Cs$_2$Cu$_{3}$SnF$_{12}$ and (b) Rb$_2$Cu$_{3}$SnF$_{12}$ viewed along the $c$ axis, where fluorine ions located outside the kagome layer are omitted. Solid and dotted lines denote the chemical unit cells. The chemical unit cell of Rb$_2$Cu$_{3}$SnF$_{12}$ is $2a\,{\times}\,2a$ enlarged. Reprinted from Ref.~\cite{Ono} and modified.}
	\label{fig:strucCsRb}
\end{figure}

Rb$_2$Cu$_3$SnF$_{12}$ has a trigonal structure ($R{\bar 3}$) at room temperature~\cite{Morita}. The unit cell in the kagome layer is enlarged to $2a\,{\times}\,2a$. Figure \ref{fig:strucCsRb} shows a comparison of the crystal structures of Cs$_2$Cu$_{3}$SnF$_{12}$ and Rb$_2$Cu$_{3}$SnF$_{12}$. The kagome lattice is modified to have four types of neighboring exchange interaction: $J_1, J_2, J_3$, and $J_4$. Figure~\ref{fig:ex_network}\,(b) shows the configuration of these exchange interactions. Pinwheels composed of exchange interactions $J_1$, $J_2$, and $J_4$ are connected by triangles of $J_3$ interactions~\cite{Morita,Matan}. The bond angles of  $J_1, J_2$, and $J_3$ are close to one another, while the bond angle of $J_4$ is significantly smaller than the others, as shown in Table~\ref{tab:exchange_bond}. Hence, the magnitudes of $J_1, J_2$, and $J_3$ are similar, and the magnitude of $J_4$ is significantly smaller than the others.

Rb$_2$Cu$_3$SnF$_{12}$ undergoes a structural phase transition at $T_{\rm t}\,{=}\,215$\,K~\cite{Matan,Downie}. The low-temperature crystal lattice was suggested to be triclinic ($P\bar{1}$)~\cite{Downie}. However, the magnetic susceptibility displays no anomaly at $T_{\rm t}$, as shown in Section 4~\cite{Morita}. Hence, the change in the exchange interaction at $T_{\rm t}$ must be tiny.

\section{Sample preparation}\label{sample}

\begin{figure}[t]
	\centering
	\includegraphics[width=15.0cm]{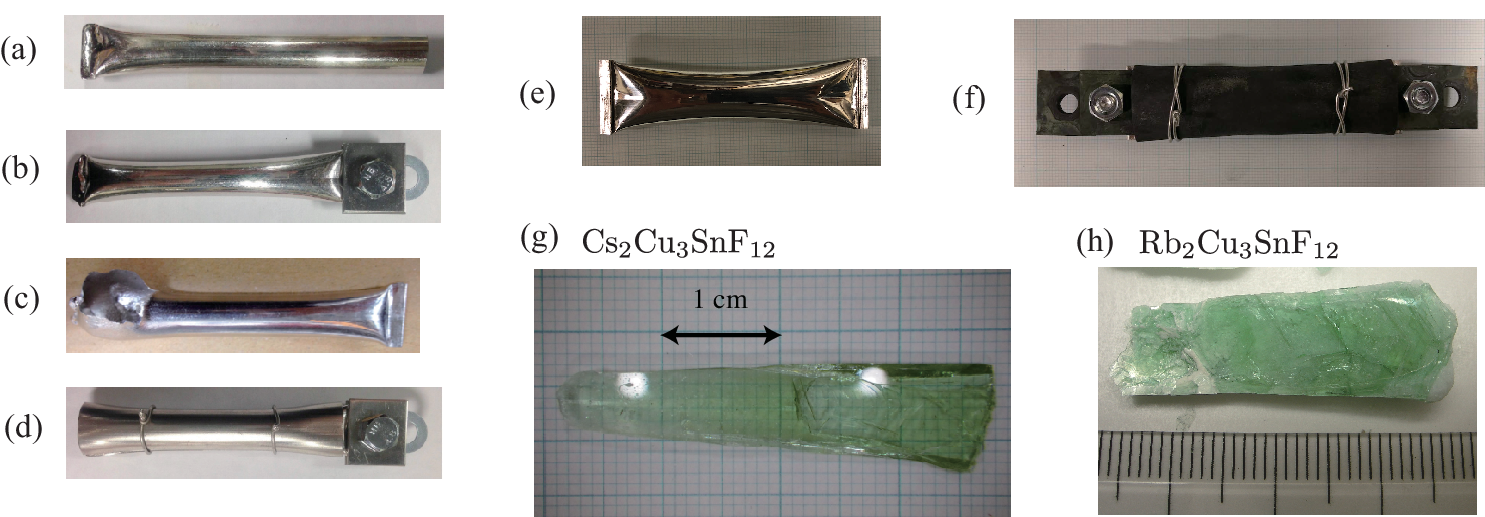}
	\caption{(a) Pt tube with an inner diameter of 10 mm. Its one end is welded. (b) Pt tube after stuffing source materials. The other end is tightly folded twice and sandwiched tightly between two Nichrome plates. (c) Bursted Pt tube due to high inner pressure. (d) Pt tube after covering with two half-pipe-shaped Nichrome protectors. Both ends of the protector are spread out by hitting with a hammer to fit the Pt tube. (e) Pt tube with an inner diameter of 15 mm after stuffing source materials. Both ends are tightly folded twice. (f) Pt tube with an inner diameter of 15 mm after sandwiching both ends tightly with two Nichrome plates and covering with two protectors. (g) and (h) Photos of Cs$_2$Cu$_{3}$SnF$_{12}$ and Rb$_2$Cu$_{3}$SnF$_{12}$ single crystals, respectively.}
	\label{fig:Pt_tube}
\end{figure}

I summarize the procedure of the crystal growth of Cs$_2$Cu$_{3}$SnF$_{12}$ and Rb$_2$Cu$_{3}$SnF$_{12}$~\cite{Ono,Morita,Saito}, which will be applicable for preparing other fluoride substances. 
Single crystals of $A_2$Cu$_3$SnF$_{12}$ with $A\,{=}\,\mathrm{Cs}$ and Rb were synthesized according to the chemical reaction 2$A$F + 3CuF$_2$ + SnF$_4 \rightarrow$ $A_2$Cu$_3$SnF$_{12}$. Pt tubes of 15 or 10 mm inner diameter and 100 mm length with 0.2 mm thickness were used for the crystal growth. One end was welded for the Pt tube with an inner diameter of 10 mm, as shown in Fig.~\ref{fig:Pt_tube}\,(a), while one end was tightly folded twice with pliers for the Pt tube with a 15 mm inner diameter (see Fig.~\ref{fig:Pt_tube}\,(e)). Source materials $A$F, CuF$_2$, and SnF$_4$ were stuffed into the Pt tubes at a ratio of approximately $3\,{:}\,3\,{:}\,2$ in a glove box filled with dry nitrogen. The Pt tube stuffed with source materials was put into a test tube and dehydrated by heating in a vacuum at $50\,{-}\,80^\circ$C for three days. The top of the test tube was stuffed with cotton wrapped in gauze, so the powdered materials were not pulled into a pump. After the dehydration, the other end of the Pt tube was tightly folded twice and sandwiched tightly between two Nichrome plates, as shown in Fig.~\ref{fig:Pt_tube}\,(b). Both ends were tightly folded twice and sandwiched tightly between two Nichrome plates for the Pt tube with a 15 mm inner diameter, as shown in Fig.~\ref{fig:Pt_tube}\,(f). We covered the Pt tubes with two half-pipe-shaped Nichrome protectors, as demonstrated in Figs.~\ref{fig:Pt_tube}\,(d) and (f) because they burst in a furnace due to high inner pressure, as shown in Fig.~\ref{fig:Pt_tube}\,(c). Single crystals were grown from the melt using a horizontal furnace. The temperature at the center of the furnace was lowered from 850 to 750$^\circ$C for Cs$_2$Cu$_3$SnF$_{12}$ and 800 to 700$^\circ$C for Rb$_2$Cu$_3$SnF$_{12}$ over 100 h. After collecting the good pieces of crystals, the same procedure was repeated using a Pt tube with an inner diameter of 13 or 10 mm and a length of 100 mm. The protectors were not utilized in this process because the inner pressure was much lower than in the first step. Transparent light-green crystals were obtained, as shown in Figs.~\ref{fig:Pt_tube}\,(g) and (h).

\section{Static magnetic properties}\label{static}

\begin{figure}[t]
	\centering
	\includegraphics[width=15.0 cm]{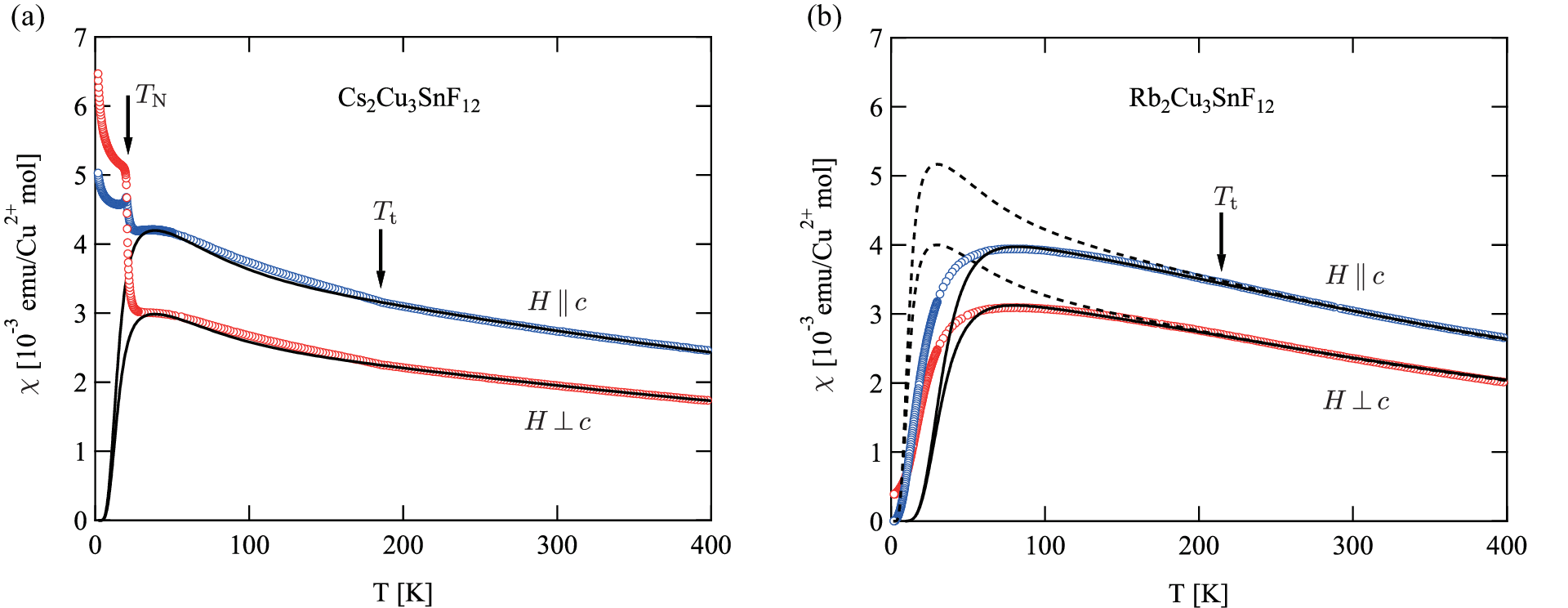}
	\caption{Temperature dependences of magnetic susceptibilities of (a) Cs$_2$Cu$_3$SnF$_{12}$ and (b) Rb$_2$Cu$_3$SnF$_{12}$ for magnetic field $H$ parallel and perpendicular to the $c$ axis~\cite{Ono,Morita}. The data of Rb$_2$Cu$_3$SnF$_{12}$ were corrected for the small Curie term due to impurities below 10 K. Arrows in (a) denote structural and magnetic phase transition temperatures $T_{\rm t}\,{=}\,185$\,K and $T_{\rm N}\,{=}\,20.0$\,K. Solid lines are the fits with $J/k_{\rm B}\,{=}\,240$\,K, $g^{\parallel}\,{=}\,2.48$, and $g^{\perp}\,{=}\,2.10$, using the theoretical susceptibilities for the spin-1/2 uniform KLHAF obtained by exact diagonalization for the 24-site kagome cluster~\cite{Misguich2}. An arrow in (b) indicates a structural phase transition temperature $T_{\rm t}\,{=}\,215$\,K. Solid lines are the results obtained by exact diagonalization for a 12-site kagome cluster with the exchange parameters and $g$-factors $J_1/k_{\rm B}\,{=}\,234$\,K, $J_2/k_{\rm B}\,{=}\,211$\,K, $J_3/k_{\rm B}\,{=}\,187$\,K and $J_4/k_{\rm B}\,{=}\,108$\,K, and $g^{\parallel}\,{=}\,2.44$ and $g^{\perp}\,{=}\,2.15$~\cite{Morita}. The dashed lines denote the fits using the theoretical susceptibilities for the spin-1/2 uniform KLHAF with $J/k_{\rm B}\,{=}\,187$\,K, $g^{\parallel}\,{=}\,2.43$, and $g^{\perp}\,{=}\,2.13$.}
	\label{fig:sus}
\end{figure}

\begin{figure}[t]
	\centering
	\includegraphics[width=15cm]{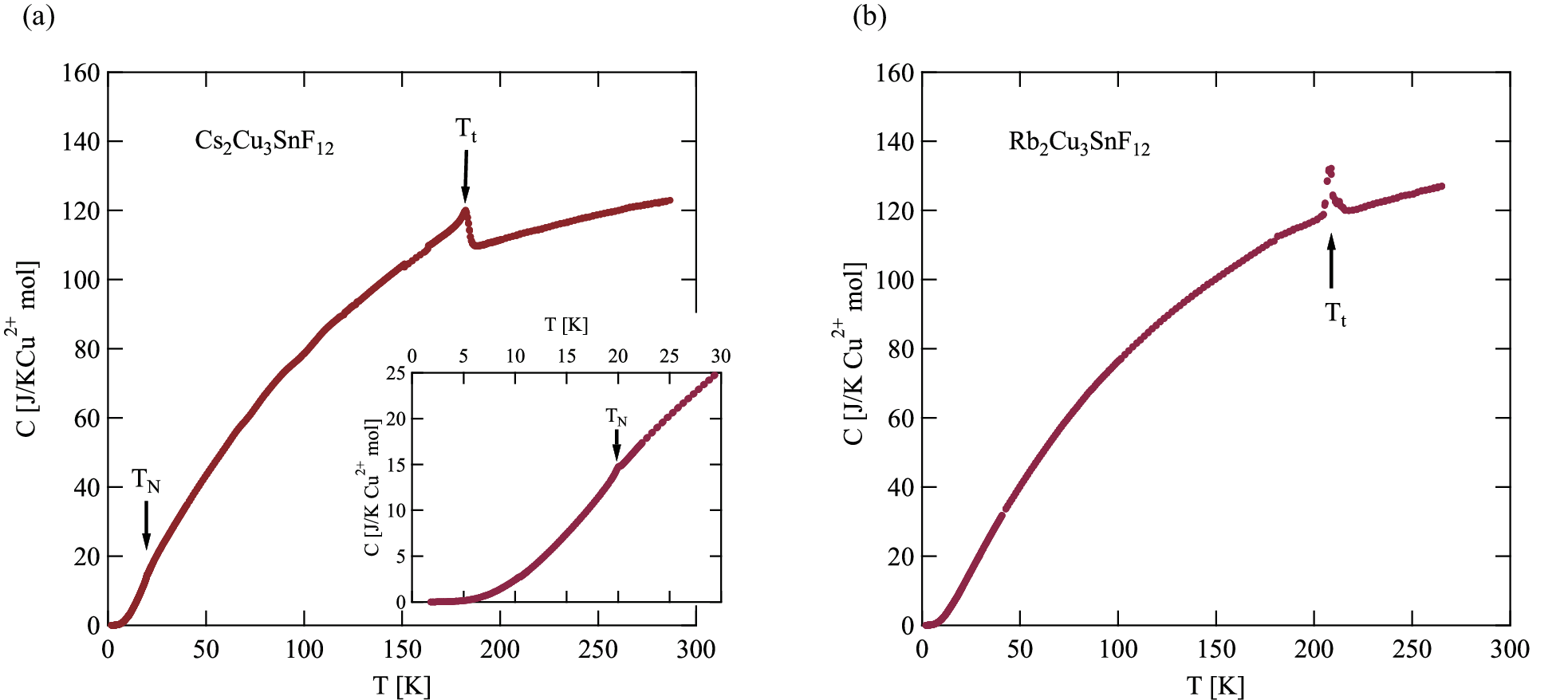}
	\caption{Specific heat as a function of temperature (a) for Cs$_2$Cu$_3$SnF$_{12}$ and (b) for Rb$_2$Cu$_3$SnF$_{12}$. Vertical arrows denote structural and magnetic phase transitions. The inset of (a) is the low-temperature specific heat for Cs$_2$Cu$_3$SnF$_{12}$.}
	\label{fig:heat}
\end{figure}

The magnetic susceptibilities in Cs$_2$Cu$_3$SnF$_{12}$ and Rb$_2$Cu$_3$SnF$_{12}$ for magnetic field parallel and perpendicular to the $c$ axis were measured using single crystals~\cite{Morita,Ono}. Figures~\ref{fig:sus} show the experimental results. 
The in-plane component $D^{\perp}$ of the $\bm D$ vector of the DM interaction is negligible in Cs$_2$Cu$_3$SnF$_{12}$, and the ratio of the out-of-plane component $D^{\parallel}$ to the exchange constant is $D^{\parallel}/J\,{\simeq}\,0.1$, as shown later. Since the effect of $D^{\parallel}$ on the magnetic susceptibility is very small for $D^{\parallel}/J\,{\simeq}\,0.1$~\cite{Rigol}, we fit a theoretical susceptibility calculated for a Heisenberg model to evaluate the exchange constant. The solid lines in Fig.~\ref{fig:sus}\,(a) denote the fits for Cs$_2$Cu$_3$SnF$_{12}$ with $J/k_{\rm B}\,{=}\,240$\,K, $g^{\parallel}\,{=}\,2.48$, and $g^{\perp}\,{=}\,2.10$, using the theoretical susceptibilities for the spin-1/2 uniform KLHAF obtained by exact diagonalization for the 24-site kagome cluster~\cite{Misguich2}. The experimental magnetic susceptibilities agree with the theory down to $T_{\rm max}\,{=}\,(1/6)J/k_{\rm B}$ displaying a rounded maximum. A slight bend anomaly at $T_{\rm t}\,{=}\,185$\,K arises from a structural phase transition. The susceptibility anomaly at $T_{\rm t}$ in Cs$_2$Cu$_3$SnF$_{12}$ contrasts with the cases of Cs$_2$Cu$_3$ZrF$_{12}$ and Cs$_2$Cu$_3$HfF$_{12}$, where magnetic susceptibilities display a significant discontinuous jump~\cite{Ono}.
The $\lambda$-like specific heat anomaly at $T_{\rm t}$ shown in Fig.~\ref{fig:heat}\,(a) indicates that this transition is close to the second-order transition. From the facts that the deviation of the crystal lattice from the room-temperature structure is small, as discussed in Section \ref{structure}, and the anomaly of magnetic susceptibility at  $T_{\rm t}$ is slight, Cs$_2$Cu$_3$SnF$_{12}$ can approximate a spin-1/2 uniform KLAF. 

Cs$_2$Cu$_3$SnF$_{12}$ undergoes a magnetic phase transition at $T_{\rm N}\,{=}\,20.0$\,K, as shown in Fig.~\ref{fig:sus}\,(a), induced by the DM interaction and finite interlayer exchange interaction~\cite{Ono}. Magnetic susceptibilities exhibit sharp anomalies at $T_{\rm N}$. The neutron diffraction experiment revealed that the spin structure in the ordered state is a ${\bm q}\,{=}\,0$ structure with positive chirality~\cite{Ono2}. 
The inset of Fig.~\ref{fig:heat}\,(a) shows the low-temperature specific heat for Cs$_2$Cu$_3$SnF$_{12}$~\cite{Katayama}. A cusp anomaly owing to magnetic ordering is observed at $T_{\rm N}\,{=}\,20.0$\,K. The tiny anomaly around $T_{\rm N}$ indicates that little entropy remains for magnetic ordering because of the well-developed short-range spin correlation caused by the exchange interaction of $J/k_{\rm B}\,{=}\,240$\,K and good two-dimensionality.

The configuration of the $\bm D$ vector of the DM interaction shown in Fig.~\ref{fig:DM}\,(a) tends to stabilize the ${\bm q}\,{=}\,0$ structure with positive chirality. 
The effect of the DM interaction on the ground state was theoretically discussed based on a model expressed as
\begin{eqnarray}
\mathcal{H}=\sum_{\langle i,j\rangle} J\left({\bm{S}}_i\cdot{\bm{S}}_j\right) + \sum_{\langle i,j\rangle} D^{\parallel}_{ij}\left(S_i^xS_j^y - S_i^yS_j^x\right).
\label{model_1}
\end{eqnarray}
With increasing $D^{\parallel}/J$, a quantum phase transition from the spin liquid state to the ${\bm q}\,{=}\,0$ state occurs~\cite{Cepas}. Although the critical value of $(D^{\parallel}/J)_{\rm c}$ was initially obtained to be approximately 0.1~\cite{Cepas,Rousochatzakis,Hering}, subsequent high-precision calculations have determined it to be $0.012\,{-}\,0.04$~\cite{Lee2,Ferrari3}. In Cs$_2$Cu$_3$SnF$_{12}$, $D^{\parallel}/J$ is evaluated as 0.12, as shown in the next Section, and exceeds the critical value $(D^{\parallel}/J)_{\rm c}\,{=}\,0.012\,{-}\,0.04$. For this reason, Cs$_2$Cu$_3$SnF$_{12}$ has an ordered ground state. In the absence of additional effects such as exchange disorder, most actual copper KLAFs will have the ordered ground state because the magnitude of the $\bm D$ vector can be estimated as $D/J\,{\simeq}\,{\Delta}g/g\,{\simeq}\,0.1$. For example, the size of $D^{\parallel}$ was evaluated as $D^{\parallel}/J\,{=}\,0.08$ and 0.25 for ZnCu$_3$(OH)$_6$Cl$_2$ and YCu$_3$(OH)$_6$Cl$_3$, respectively~\cite{Zorko2,Arh}, which exceed the critical value.

Figure~\ref{fig:sus}\,(b) shows magnetic susceptibilities in Rb$_2$Cu$_3$SnF$_{12}$~\cite{Morita,Ono}. The data were corrected for the small Curie term due to impurities. The raw susceptibility data are shown in Ref.~\cite{Morita}. With decreasing temperature, the magnetic susceptibilities have broad maxima near 80\,K and decrease towards zero. This indicates that the ground state is a singlet with an excitation gap. The magnitude of the gap was evaluated to be ${\Delta}/k_{\rm B}\,{=}\,27.3$\,K by INS experiment~\cite{Matan,Matan2}. Rb$_2$Cu$_3$SnF$_{12}$ undergoes a structural phase transition at $T_{\rm t}\,{=}\,215$\,K~\cite{Matan,Downie}. The specific heat measurement also observes this structural transition, as shown in Fig.~\ref{fig:heat}\,(b). However, the exchange interaction change should be little because no anomaly is observed in the magnetic susceptibility at  $T_{\rm t}\,{=}\,215$\,K.

The magnetic susceptibility of Rb$_2$Cu$_3$SnF$_{12}$ was analyzed based on the exchange network shown in Fig.~\ref{fig:ex_network}\,(b). It was found that the observed temperature dependence of the magnetic susceptibility can be reproduced only when $J_1\,{\simeq}\,J_2\,{\simeq}\,J_3\,{>}\,J_4$~\cite{Morita,Khatami}. Solid lines in Fig.~\ref{fig:sus}\,(b) are the results obtained by exact diagonalization for a 12-site kagome cluster with the exchange parameters and $g$-factors $J_1/k_{\rm B}\,{=}\,234$\,K, $J_2\,{=}\,0.90J_1$, $J_3\,{=}\,0.80J_1$ and $J_4\,{=}\,0.46J_1$, and $g^{\parallel}\,{=}\,2.44$ and $g^{\perp}\,{=}\,2.15$~\cite{Morita}. The fits using the theoretical susceptibilities for the spin-1/2 uniform KLHAF with $J/k_{\rm B}\,{=}\,187$\,K, $g^{\parallel}\,{=}\,2.43$, and $g^{\perp}\,{=}\,2.13$ are also shown by dashed lines for comparison in Fig.~\ref{fig:sus}\,(b). These exchange parameters are consistent with those obtained from the dispersion curves of singlet-triplet excitations, as shown in the next Section. Below $T\,{=}\,70$\,K, the calculated susceptibility decreases more rapidly than the experimental susceptibility. This is ascribed to the finite-size effect. The magnitude of $J_4$ is half of the others. This indicates that only a $10\,{-}\,15^{\circ}$ difference in the bond angle drastically changes the exchange interaction when the bond angle is between 120 and 150$^{\circ}$ (see Table~\ref{tab:exchange_bond} and Fig.~\ref{fig:exchange}). This fact will hint at understanding the Cu$^{2+}$ substitution for Zn$^{2+}$ and exchange randomness in Herbertsmithite. 

Grbi\'{c} {\it et al.}~\cite{Grbic} investigated the excitation gap ${\Delta}(H)$ in the magnetic field for $H\,{\parallel}\,c$ by the NMR experiment. With increasing the magnetic field, the gap decreases. However, the gap reaches the minimum of approximately ${\Delta}(0)/2$ near $H\,{=}\,15$\,T and turns into an increase. The gap behavior was ascribed to the staggered field arising from the staggered tilt of the principal axis of the ${\bm g}$ tensor caused by the crystal structure. 

\begin{figure}[t]
	\centering
	\includegraphics[width=8cm]{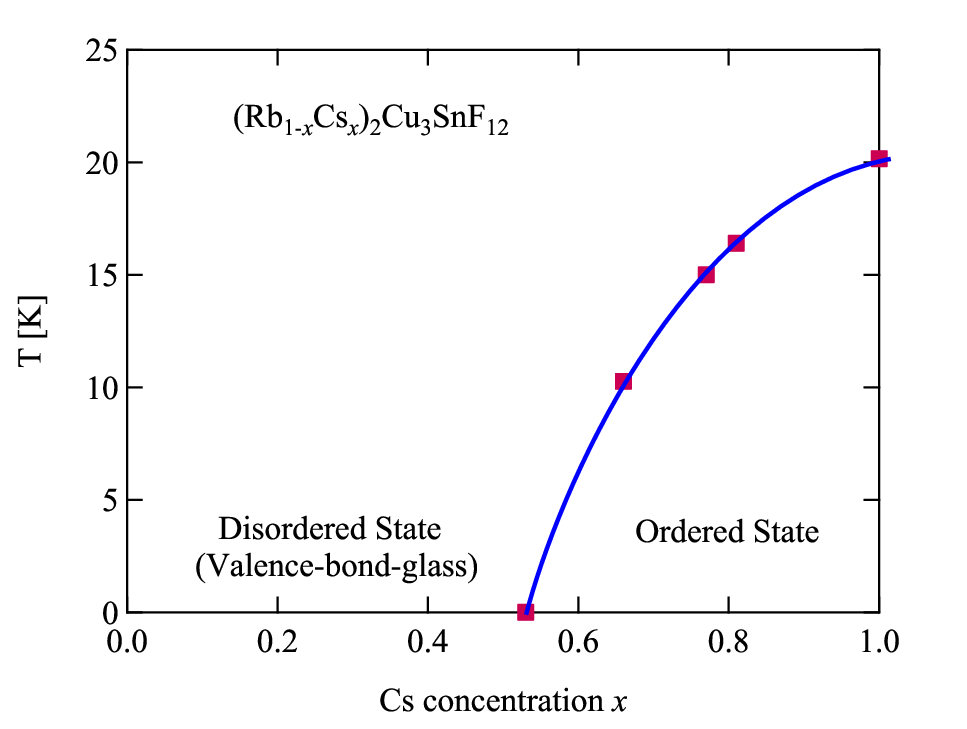}
	\caption{Temperature vs. cesium concentration $x$ phase diagram in a mixed system (Rb$_{1{-}x}$Cs$_x$)$_2$Cu$_3$SnF$_{12}$~\cite{Katayama}. The solid line is a guide for the eyes.}
	\label{fig:phase_CsRb}
\end{figure}

The magnetic ground state in a mixed system (Rb$_{1{-}x}$Cs$_x$)$_2$Cu$_3$SnF$_{12}$ was investigated by magnetization measurements and $\mu$SR experiments~\cite{Katayama,Suzuki}. Figure~\ref{fig:phase_CsRb} shows the temperature vs. cesium concentration $x$ phase diagram~\cite{Katayama}. As cesium concentration $x$ decreases from $x\,{=}\,1$, the transition temperature $T_{\rm N}$ decreases, and the ordered state vanishes at $x\,{=}\,0.53$. The $\mu$SR experiment observed the absence of magnetic ordering at $x\,{=}\,0.53$~\cite{Suzuki}. Magnetic ordering is absent for $0\,{\leq}\,x\,{\leq}\,0.53$. The ground state for $0\,{<}\,x\,{\leq}\,0.53$ is deduced to be the valence-bond-glass (VBG) state because the magnetic susceptibility is finite in this $x$ range despite the absence of magnetic ordering~\cite{Katayama}.

\section{Magnetic excitations}\label{excitations}

\begin{figure}[t]
	\centering
	\includegraphics[width=16cm]{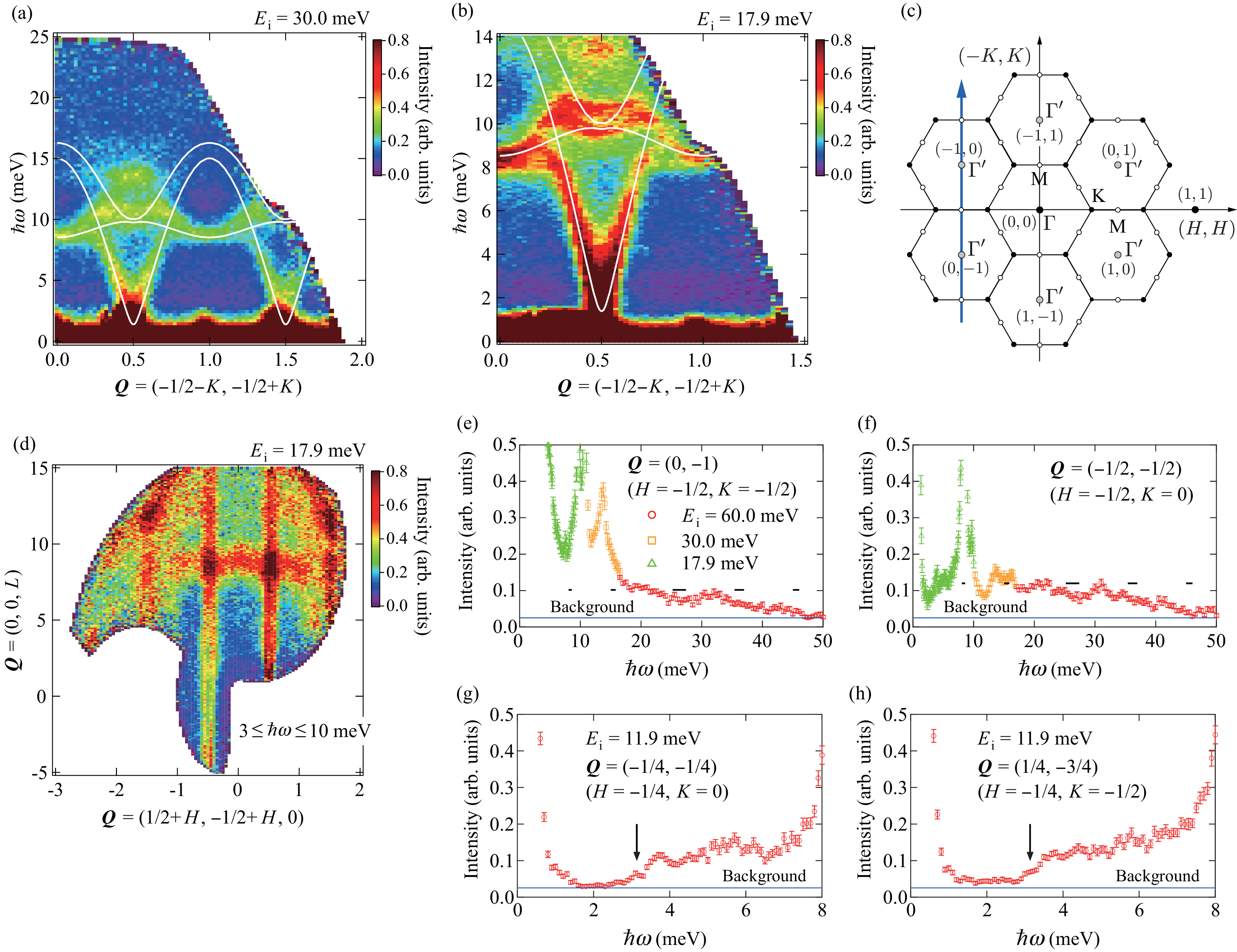}
	\caption{The excitation spectra of Cs$_2$Cu$_3$SnF$_{12}$ measured at $T\,{=}\,5$ K $({\ll}\,T_{\rm N})$~\cite{Saito}. Energy-momentum maps of scattering intensity measured with incident neutron energies of (a) $E_{\rm i}\,{=}\,30.0$ and (b) 17.9 meV along a high-symmetry direction $\bm{Q}\,{=}\,({-}1/2{-}K, {-}1/2{+}K)$ (blue line in (c)). The scattering intensities were averaged along $L$ to map the scattering intensity in the two-dimensional (2D) reciprocal lattice shown in (c), assuming good two-dimensionality. The solid lines in (a) and (b) are dispersion curves calculated by LSWT with $J\,{=}\,12.8$ meV, $J'\,{=}\,-0.043J$, $D^\parallel\,{=}\,0.18J$, and $D^\perp\,{=}\,0.062J$. (d) Scattering intensity map in the $(1/2{+}H, {-}1/2{+}H, L)$ plane measured with $E_{\mathrm{i}}\,{=}\,17.9$ meV. The energy was averaged in a range of $3\,{\leq}\,{\hbar}{\omega}\,{\leq}\,10$\,meV. (e) and (f) Scattering intensities as a function of energy measured for $\bm{Q}\,{=}\,(0, {-}1)$ and $({-}1/2, {-}1/2)$. Horizontal bars are energy resolution.
(g) and (h) Scattering intensities as a function of energy measured with incident neutron energy $E_{\mathrm{i}}\,{=}\,11.9$ meV for $\bm{Q}\,{=}\,({-}1/4, {-}1/4)$ and $(1/4, {-}3/4)$. Reprinted from Ref.~\cite{Saito} and modified.}
	\label{fig:spectra_CCSF}
\end{figure}

INS experiments are essential for understanding the nature of the ground state through magnetic excitations. The magnetic excitations of Cs$_2$Cu$_3$SnF$_{12}$ and Rb$_2$Cu$_3$SnF$_{12}$ were investigated by INS experiments~\cite{Ono2,Matan,Matan2,Saito,Matan4}. Here, I introduce the experimental results obtained by Saito {\it et al.}~\cite{Saito} and related theories. 

Figures~\ref{fig:spectra_CCSF}\,(a) and (b) show excitation spectra of Cs$_2$Cu$_3$SnF$_{12}$ measured along $\bm{Q}\,{=}\,({-}\,1/2\,{-}\,K, {-}1/2\,{+}\,K)$ (blue line in (c)) with incident neutron energies of $E_{\rm i}\,{=}\,30.0$ and 17.9\,meV~\cite{Saito}. Four strong excitations ascribed to single-magnon excitations are observed around the ${\Gamma}^{\prime}$ point. Their energies at the ${\Gamma}^{\prime}$ point are $\hbar\omega\,{=}\,1.0,\, 9.6,\, 10.7$, and 13.8 meV. Figure~\ref{fig:spectra_CCSF}\,(d) shows the scattering intensity map in the $(1/2\,{+}\,H, -1/2\,{+}\,H, L)$ plane. The scattering streaks along the $L$ direction indicate the absence of dispersion along the $c^*$ direction, i.e., good two-dimensionality.

Figures~\ref{fig:spectra_CCSF}\,(e) and (f) show scattering intensities as a function of energy measured for $\bm{Q}\,{=}\,(0, -1)$ and $(-1/2, -1/2)$. The excitation continuum extends up to approximately 50 meV, nearly equal to $2.5J$. The spectral weight of the excitation continuum is considerably larger than that of the single-magnon excitations, as observed in the spin-1/2 triangular-lattice Heisenberg-like antiferromagnet Ba$_3$CoSb$_2$O$_9$~\cite{Ito,Macdougal}. As seen from Fig.~\ref{fig:spectra_CCSF}\,(b), the scattering intensity changes at approximately 3 meV irrespective of the momentum transfer $\bm{Q}$ except in the vicinities of the ${\Gamma}^{\prime}$ points. Figures~\ref{fig:spectra_CCSF}\,(g) and (h) show the scattering intensities as a function of energy for $\bm{Q}\,{=}\,(-1/4, -1/4)$ and $(1/4, -3/4)$. The scattering intensity increases rapidly at approximately $\hbar\omega\,{=}\,3$\,meV and increases gradually with increasing excitation energy. These results show that the broad excitation continuum without a marked structure spreads in a wide energy range from 3 meV ($0.15J$) to approximately 50 meV ($2.5J$). Recent theories based on the fermionic approach of spinon excitations from the spin liquid ground state demonstrated that a broad excitation continuum extends to $2.7J$~\cite{Zhang2,Ferrari2}. The upper bound energy of the excitation continuum is consistent with $2.5J$ observed in Cs$_2$Cu$_3$SnF$_{12}$. 

The dispersion curves of single-magnon excitations were analyzed first by the linear spin wave theory (LSWT) based on a model of 
\begin{eqnarray}
\mathcal{H}=\sum_{\langle i,j\rangle} J\left({\bm{S}}_i\cdot{\bm{S}}_j\right) + \sum_{\langle i,j'\rangle} J'\left({\bm{S}}_i\cdot{\bm{S}}_{j'}\right) + \sum_{\langle i,j\rangle} {\bm{D}}_{ij}\cdot\left({\bm{S}}_i\times{\bm{S}}_j\right),
\label{model_2}
\end{eqnarray}
where the first and second terms are the NN and next-nearest-neighbor (NNN) exchange interactions, respectively. The third term is the DM interaction between the nearest-neighbor spins. The white solid lines in Figs.~\ref{fig:spectra_CCSF}\,(a) and (b) are dispersion curves calculated using LSWT with $J\,{=}\,12.8$ meV, $J'\,{=}\,-0.043J$, $D^{\parallel}\,{=}\,0.18J$, and $D^\perp\,{=}\,0.062J$~\cite{Saito}. Almost the same parameters were obtained in Ref.~\cite{Matan4}. As for the DM interaction, the out-of-plane component $D^{\parallel}$ is much larger than the in-plane component $D^{\perp}$. However, $D^\perp$ is necessary to produce a small gap of ${\Delta}\,{=}\,1.0$\,meV at the ${\Gamma}^{\prime}$ points and a small split of two branches near ${\hbar}{\omega}\,{=}\,10$\,meV at the ${\Gamma}^{\prime}$ points.

The slope of the linear dispersion in the vicinity of the ${\Gamma}^{\prime}$ points determines the NN exchange constant. Although low-energy dispersion curves for ${\hbar}{\omega}\,{\leq}\,10$\,meV appear to be well reproduced by the LSWT calculation, the obtained NN exchange constant $J_{\rm disp}\,{=}\,12.8$ meV is significantly smaller than $J_{\rm{mag}}\,{=}\,20.7$ meV obtained from the analysis of the temperature dependence of magnetic susceptibility~\cite{Ono}. This indicates that the excitation energy of the single magnon is largely renormalized downward by the quantum many-body effect because the $J_{\rm{mag}}$ should correspond to the intrinsic exchange constant.  

For high-energy single-magnon excitations above ${\hbar}{\omega}\,{=}\,11$\,meV, there is a significant difference between the experimental and LSWT results. High-energy single-magnon excitations near the zone boundary expected at ${\hbar}{\omega}\,{\simeq}\,15$\,meV from LSWT are not observed. The fourth intense single-magnon excitation of ${\hbar}{\omega}\,{\simeq}\,14$ meV near the ${\Gamma}^{\prime}$ point cannot be explained by LSWT because it derives only three transverse modes. Recently, the magnetic excitations in spin-1/2 TLHAF were discussed based on spinon excitations from the QSL state superposed with the ordered state~\cite{Zhang,Ghioldi3} and the tensor network approach~\cite{Chi}. In spinon theories, single-magnon excitations are described as the bound spinons. These theories all predict three low-energy transverse modes and one intermediate-energy damped longitudinal mode and explain most of the characteristics of single-magnon excitations observed in Ba$_3$CoSb$_2$O$_9$~\cite{Ito,Macdougal}. 
Referring to these theories, the fourth intense mode observed in Cs$_2$Cu$_3$SnF$_{12}$ was deduced to be longitudinal, i.e., amplitude mode~\cite{Saito}. However, the mode will not be a pure longitudinal mode but a hybrid mode of the transverse and longitudinal modes, as observed in CsFeCl$_3$~\cite{Hayashida}, because the ordered spin structure is non-collinear. 

\begin{figure}[t]
	\centering
	\includegraphics[width=8cm]{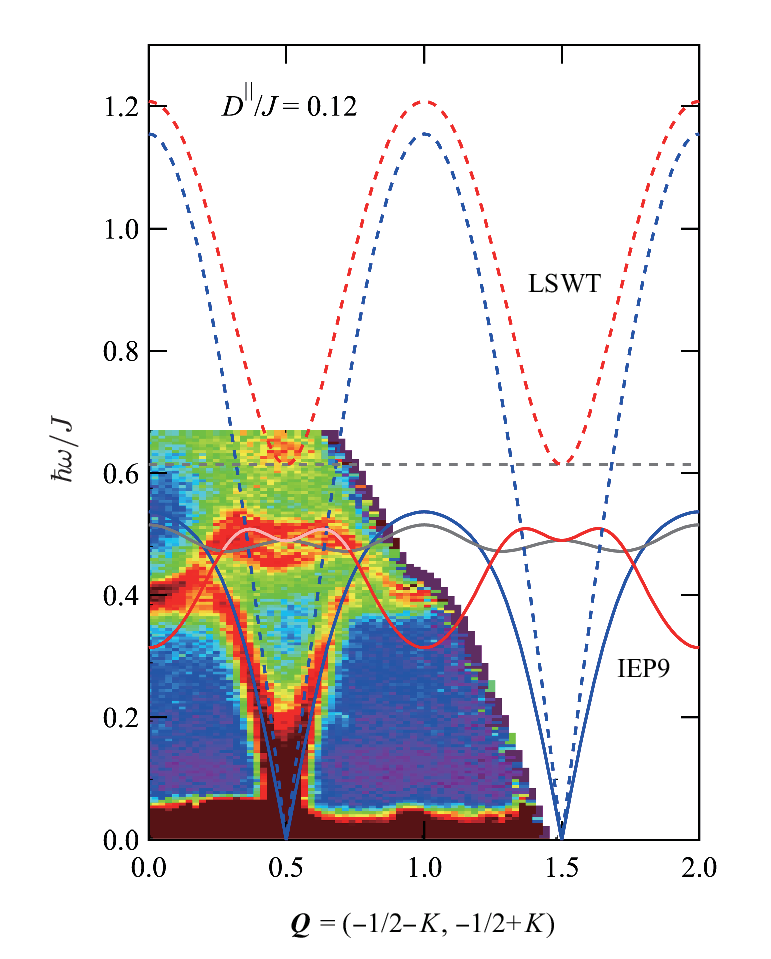}
	\caption{Dispersion curves along $\bm{Q}\,{=}\,({-}1/2{-}K, {-}1/2{+}K)$ computed by the ninth-order IPE (solid lines) and LSWT (dashed lines) with $J\,{=}\,20.7$\,meV and $D^{\parallel}/J\,{=}\,0.12$~\cite{Kogure}. The color map is the excitation spectrum of Cs$_2$Cu$_3$SnF$_{12}$. Reprinted from Ref.~\cite{Kogure} and modified ({\copyright} 2023 The Physical Society of Japan).}
	\label{fig:SE_CCSF}
\end{figure}

Based on a minimal model expressed by Eq.~(\ref{model_1}), Ferrari {\it et al.}~\cite{Ferrari2} discussed the excitation spectrum of the spin-1/2 KLAF using a variational Monte Carlo approach combined with a fermionic spinon theory. Their results capture the characteristics of the excitation spectrum of Cs$_2$Cu$_3$SnF$_{12}$, i.e., the dispersion curves of single-magnon modes and a broad excitation continuum. However, they used the renormalized exchange constant $J\,{=}\,12.8$\,meV, not the intrinsic $J\,{=}\,20.7$\,meV, and $D^{\parallel}/J\,{=}\,0.18$ obtained by the fit using LSWT~\cite{Saito}.

Applying the Ising expansion (IEP) method to the minimal model expressed by Eq.~(\ref{model_1}), Kogure {\it et al.}~\cite{Kogure} calculated the dispersion curves for three transverse single-magnon modes. The IEP method was successfully applied to compute the transverse single-magnon modes for the spin-1/2 TLHAF~\cite{Zheng}. Three solid lines in Fig.~\ref{fig:SE_CCSF} denote the dispersion curves calculated by the ninth-order IPE with $J\,{=}\,20.7$\,meV and $D^{\parallel}/J\,{=}\,0.12$. Although there are certain discrepancies around the M point, the IEP result successfully describes the experimental dispersion curves around the ${\Gamma}^{\prime}$ point. The value of $D^{\parallel}/J\,{=}\,0.12$ appears more reasonable than $D^{\parallel}/J\,{=}\,0.18$ obtained by applying LSWT.
The dashed lines in Fig.~\ref{fig:SE_CCSF} are the LSWT results calculated with the same parameters. The excitation energies of the LSWT are drastically renormalized downwards due to the quantum many-body effect. For the mode depicted in red, the dispersion mountain at the M point for LSWT changes into the valley. This phenomenon is similar to the roton-like local minimum in the dispersion curve for the spin-1/2 TLHAF~\cite{Zheng,Ito,Macdougal}, but its manifestation is even more dramatic. Hence, it will be a prototypical example of the avoided quasiparticle decay~\cite{Verresen}. 
Another notable point is that the flat mode for the LSWT acquires dispersion without the help of the NNN interaction. The behavior of the single-magnon excitations observed in Cs$_2$Cu$_3$SnF$_{12}$ contrasts sharply with that in a classical KLAF Fe-based Jarosite KFe$_3$(OH)$_6$(SO$_4$)$_2$~\cite{Matan5}, for which the LSWT almost perfectly describes its dispersion curves~\cite{Yildirim}.

The intense broad excitation continuum extending from $0.15J$ to approximately $2.5J$ observed in Cs$_2$Cu$_3$SnF$_{12}$ is consistent with the results of fermionic spinon theories~\cite{Zhang2,Ferrari2}. The dispersion curves of single magnon excitations in Cs$_2$Cu$_3$SnF$_{12}$ are much different from the LSWT result, and the fermionic spinon theory and the IEP describe their main characteristics.
These facts strongly suggest that the excitation continuum in Cs$_2$Cu$_3$SnF$_{12}$ is the spinon continuum and that the single-magnon excitations are bound states of spinons. Because spinons are characteristic excitations of the QSL, it can be deduced that a sizable QSL component coexists with an ordered component in the ground state in Cs$_2$Cu$_3$SnF$_{12}$. The following fact also supports this. In a quasi-1D spin 1/2 Heisenberg antiferromagnet KCuF$_3$, an excitation spectrum identical to that of an isolated spin-1/2 Heisenberg antiferromagnetic chain was observed by the INS experiment even below the ordering temperature $T_{\rm N}\,{=}\,39$\,K~\cite{Lake}. A two-spinon continuum with the lower bound given by the des Cloizeaux-Pearson mode was confirmed as predicted by theories~\cite{dCP,Faddeev,Muller2}. This indicates that a significant spin fluid component is superposed to the ordered component in the ground state of KCuF$_3$.
 
\begin{figure}[t]
	\centering
	\includegraphics[width=16cm]{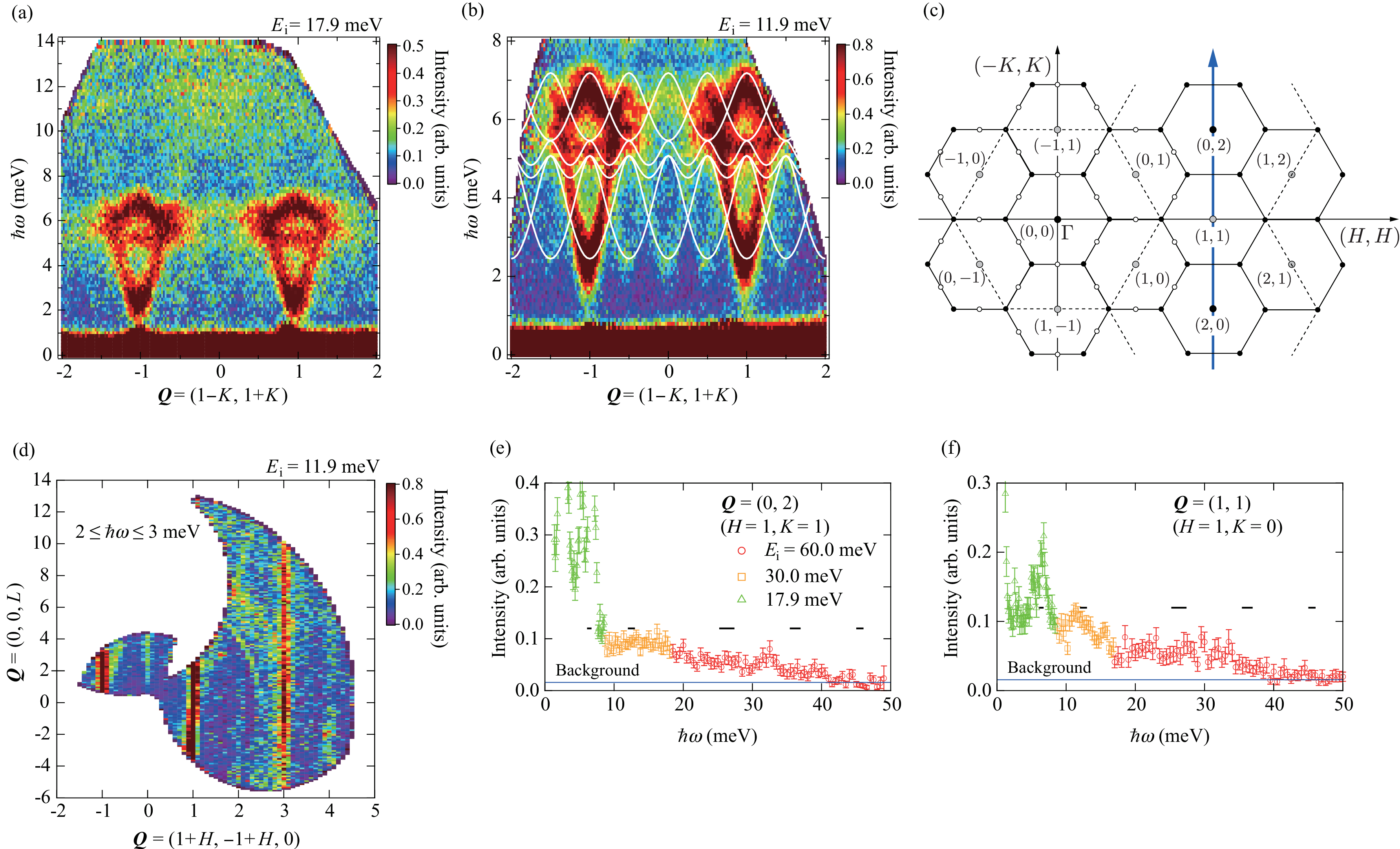}
	\caption{The excitation spectra of Rb$_2$Cu$_3$SnF$_{12}$ measured at $T\,{=}\,5$ K~\cite{Saito}. Energy-momentum maps of scattering intensity measured with incident neutron energies of (a) $E_{\rm i}\,{=}\,17.9$ and (b) 11.9\,meV along a high-symmetry direction $\bm{Q}\,{=}\,(1{-}K, 1{+}K)$ indicated by blue line in the 2D reciprocal lattice shown in (c). The scattering intensities were averaged along $L$ to map the scattering intensity in the 2D reciprocal lattice, assuming good two-dimensionality. The solid and dashed lines in (c) are elementary BZ boundaries for the $2a\,{\times}\,2a$ enlarged chemical unit cell at room temperature of Rb$_2$Cu$_3$SnF$_{12}$ and the uniform kagome lattice, respectively. The solid lines in (b) are calculated dispersion curves for main singlet-triplet excitations and their ghost modes. (d) Scattering intensity map in the $(1{+}H, {-}1{+}H, L)$ plane measured with $E_{\mathrm{i}}\,{=}\,11.9$ meV. The energy was averaged in a range of $2\,{\leq}\,{\hbar}{\omega}\,{\leq}\,3$\,meV. (e) and (f) Scattering intensities as a function of energy measured for $\bm{Q}\,{=}\,(0, 2)$ and $(1, 1)$. Horizontal bars are energy resolution. Reprinted from Ref.~\cite{Saito} and modified.}
	\label{fig:spectra_RCSF}
\end{figure}

Figures~\ref{fig:spectra_RCSF}\,(a) and (b) show energy-momentum maps of scattering intensity in Rb$_2$Cu$_3$SnF$_{12}$ measured with incident neutron energies of (a) $E_{\rm i}\,{=}\,17.9$ and (b) 11.9\,meV along a high-symmetry direction $\bm{Q}\,{=}\,(1{-}K, 1{+}K)$~\cite{Saito}.
The excitation data were analyzed based on the $2a\,{\times}\,2a$ enlarged chemical unit cell at room temperature~\cite{Saito}. The solid and dashed lines in Fig.~\ref{fig:spectra_RCSF}\,(c) are elementary Brillouin zone (BZ) boundaries for the $2a\,{\times}\,2a$ enlarged chemical unit cell and the uniform kagome lattice, respectively. Strong excitations centered at $\bm{Q}\,{=}\,(2, 0)$ and $(0, 2)$ are the triplet excitations from the singlet ground state. Three weak excitations between the main singlet-triplet excitations are their ghost modes originating from the contraction of the BZs caused by the enlargements of the chemical unit cell at room temperature and low temperatures below the structural phase transition at $T_{\rm t}\,{=}\,215$\,K~\cite{Matan,Downie}. Figure~\ref{fig:spectra_RCSF}\,(d) shows the scattering intensity map in the $(1\,{+}\,H, {-}1\,{+}\,H, L)$ plane for the energy range of $2\,{\leq}\,{\hbar}{\omega}\,{\leq}\,3$\,meV. Strong scattering streaks for the singlet-triplet excitations at $H\,{=}\,2n+1$ indicate good two-dimensionality.

Within the Heisenberg model, the lowest-energy singlet-triplet excitation is expected to be located at the K point given by $\bm{Q}\,{=}\,(1/3, 1/3)$ and equivalent points~\cite{Yang2}. However, the lowest excitation occurs at the $\Gamma$ point, as shown in Figs~\ref{fig:spectra_RCSF}\,(a) and (b). This is because the out-of-plane component $D_{ij}^\parallel$ of the DM interaction splits the triply degenerate triplet modes into $S^z\,{=}\,0$ and $S^z\,{=}\,{\pm}1$ modes, and with increasing magnitude of $D_{ij}^\parallel$, the position of the lowest energy in the $S^z\,{=}\,{\pm}1$ mode changes from the K point to the  $\Gamma$ point~\cite{Matan,Hwang2}. The dispersion curve for the $S^z\,{=}\,0$ mode is the same as that for the pure Heisenberg model because the out-of-plane component $D_{ij}^\parallel$ does not influence the $S^z\,{=}\,0$ mode.

The singlet-triplet excitations were analyzed using the dimer series expansion up to eighth order based on a model expressed as~\cite{Matan,Matan2}
\begin{equation}
\mathcal{H}=\sum_{\langle i,j\rangle} J_{ij}\left(\bm{S}_i\cdot\bm{S}_j\right) + \sum_{\langle i,j\rangle} D^{\parallel}_{ij}\left(S_i^xS_j^y - S_i^yS_j^x\right),
\label{model_RCSF}
\end{equation} 
where the configuration of four NN exchange interactions and the DM interactions are shown in Figs.~\ref{fig:DM} and \ref{fig:ex_network}. Here, the in-plane component $D_\alpha^\perp$ of the $\bm{D}$ vectors was neglected for simplification. Singlet dimers are assumed to be located on the strongest exchange bond $J_1$ to form a pinwheel VBS. The best fit can be obtained with $J_1\,{=}\,18.6$ meV, $J_2\,{=}\,0.95J_1$, $J_3\,{=}\,0.85J_1$, $J_4\,{=}\,0.55J_1$, and $D_\alpha^\parallel\,{=}\,0.18J_\alpha$ $({\alpha}\,{=}1, 2, 3$, and 4)~\cite{Matan,Matan2}. Below $T_{\rm t}\,{\,{=}\,}\,215$\,K, the $4a\,{\times}\,4a$ enlarged chemical unit cell was assumed to approximate the crystal structure. The magnitudes of $J_{ij}$ and $D_\alpha^\parallel$ are slightly modified due to the enlargement of the chemical unit cell. This perturbation gives rise to the ghost modes. The solid lines in Fig.~\ref{fig:spectra_RCSF}\,(b) are the dispersion curves for the singlet-triplet excitations and their ghost modes. The exchange parameters obtained by this analysis are consistent with those evaluated from the magnetic susceptibility data~\cite{Morita,Khatami}.

As shown in Fig.~\ref{fig:spectra_RCSF}\,(a), the structured excitation continuum is observed in the energy range of $7\,{\leq}\,{\hbar}{\omega}\,{\leq}\,14$ meV, which is just above the energy range of the singlet-triplet excitation ($2\,{\leq}\,{\hbar}{\omega}\,{\leq}\,7$\,meV). The scattering intensity has a broad maximum at ${\hbar}{\omega}\,{\simeq}\,12$ meV, which is approximately twice the energy of the upper branch of the singlet-triplet excitation. Hence, the structured excitation continuum for $7\,{\leq}\,{\hbar}{\omega}\,{\leq}\,14$ meV was attributed to two-triplet excitations~\cite{Saito}.

 Figures~\ref{fig:spectra_RCSF}\,(e) and (f) show scattering intensities as a function of energy measured for $\bm{Q}\,{=}\,(0, 2)$ and $(1,1)$. A broad excitation continuum extends above the singlet-triplet and two-triplet excitations. The upper bound of the excitation continuum is approximately 40\,meV, which is 2.6 times larger than the average of the four NN exchange interactions $J_{\rm{avg}}\,{=}\,15.6$\,meV. This upper bound is almost the same as that observed in Cs$_2$Cu$_3$SnF$_{12}$ and consistent with the theoretical upper bound of $2.7J$ obtained by a fermionic approach of spinon excitations from the spin liquid ground state~\cite{Zhang2,Ferrari2}. The weight of the excitation continuum is larger than the weight of the singlet-triplet excitation,  as observed in Cs$_2$Cu$_3$SnF$_{12}$. These observations indicate that the high-energy excitation continuum in spin-1/2 Heisenberg-like KLAFs has common characteristics, irrespective of the ground state.

\section{Conclusion}
I reviewed the crystal structures and static and dynamic properties of spin-1/2 KLAFs Cs$_2$Cu$_3$SnF$_{12}$ and Rb$_2$Cu$_3$SnF$_{12}$. I described in detail the procedure of crystal growth. This will be applicable for preparing other fluoride materials.
I also discussed that the exchange network in Cs$_2$Cu$_3$SnF$_{12}$ is approximately uniform from the viewpoint of the bond angle. The following was found from the experiments on Cs$_2$Cu$_3$SnF$_{12}$ and Rb$_2$Cu$_3$SnF$_{12}$ and related theoretical studies. Cs$_2$Cu$_3$SnF$_{12}$ can approximate a spin-1/2 uniform KLAF described by the model of Eq.~(\ref{model_1}) with $J\,{=}\,20.7$\,meV and $D^{\parallel}/J\,{=}\,0.12$. The DM interaction gives rise to an ordered ${\bm q}\,{=}\,0$ ground state with positive chirality. The magnetic excitation spectrum consists of single-magnon excitations and an intense broad excitation continuum ranging from $0.15J$ to approximately $2.5J$. The energies of the single-magnon modes are drastically renormalized downwards owing to a remarkable quantum many-body effect. These dynamic properties indicate that the ground state has a sizable QSL component superposed to the ordered component. The ground state of Rb$_2$Cu$_3$SnF$_{12}$ is a pinwheel VBS owing to the exchange network shown in Fig.~\ref{fig:ex_network}\,(b). Magnetic susceptibility data confirmed the gapped singlet ground of Rb$_2$Cu$_3$SnF$_{12}$. The dispersion curves of singlet-triplet excitations measured by INS experiments were successfully analyzed using a model expressed by Eq.~(\ref{model_RCSF}), and the interaction parameters were determined. High-energy excitations comprise two-triplet excitations and an intense broad excitation continuum with the upper bound of $2.6J_{\rm avg}$. The characteristics of the broad high-energy excitation continuum are common to spin-1/2 Heisenberg-like KLAFs, irrespective of the ground states.

\section*{Acknowledgments}
The author thanks K. Matan, K. Morita, and Y. Fukumoto for stimulating and fruitful discussions.


\section*{References}

\begin{thebibliography}{99} 

\bibitem{Balents_2010} Balents L 2010 {\it Nature (London)} \textbf{464} 199.

\bibitem{Broholm} Broholm C, Cava R J, Kivelson S A, Nocera D G, Norman M R and Senthil T 2020 Quantum spin liquids {\it Science} \textbf{367} 263.

\bibitem{Anderson_1973} Anderson P W 1973 {\it Mater. Res. Bull.} \textbf{8} 153.

\bibitem{Miyashita_1984} Miyashita S 1984 {\it J. Phys. Sic. Jpn.} \textbf{53} 44.

\bibitem{Huse} Huse D A and Elser V 1988 {\it Phys. Rev. Lett.} \textbf{60} 2531.

\bibitem{Jolicoeur} Jolicoeur Th and Le Guillou J C 1989 {\it Phys. Rev. B} \textbf{40} 2727(R).

\bibitem{Bernu} Bernu B, Lecheminant P, Lhuillier C and Pierre L 1994 {\it Phys. Rev. B} \textbf{50} 10048.

\bibitem{Singh} Singh R R P and Huse D A 1992 {\it Phys. Rev. Lett.} \textbf{68} 1766.

\bibitem{White} White S R and Chernyshev A L 2007 {\it Phys. Rev. Lett.} \textbf{99} 127004.

\bibitem{Gotze} G\"{o}tze O, Richter J, Zinke R and Farnell D J J 2016 {\it J. Magn. Magn. Mater.} \textbf{397} 333.

\bibitem{Li2} Li Q, Li H, Zhao J, Luo H G and Xie Z Y 2022 {\it Phys. Rev. B}  \textbf{105} 184418.

\bibitem{Nishimori} Nishimori H and Miyashita S 1986 {\it J. Phys. Soc. Jpn.} \textbf{55} 4448.

\bibitem{Chubokov} Chubokov A V and Golosov D I 1991 {\it J. Phys.: Condens. Matter} \textbf{3} 69.

\bibitem{Nikuni} Nikuni T and Shiba H 1993 {\it J. Phys. Soc. Jpn.} \textbf{62} 3268.

\bibitem{Honecker} Honecker A 1999 {\it J. Phys.: Condens. Matter} \textbf{11} 4697.

\bibitem{Alicea} Alicea J, Chubukov A V and Starykh O A 2009 {\it Phys. Rev. Lett.} \textbf{102} 137201.

\bibitem{Farnell} Farnell D J J, Zinke R, Schulenburg J and Richter J 2009 {\it J. Phys.: Condens. Matter} \textbf{21} 406002.

\bibitem{Sakai} Sakai T and Nakano H  2011 {\it Phys. Rev. B} \textbf{83} 100405(R).

\bibitem{Hotta} Hotta C, Nishimoto S and Shibata N 2013 {\it Phys. Rev. B} \textbf{87} 115128.

\bibitem{Yamamoto1} Yamamoto D, Marmorini G and Danshita I 2014 {\it Phys. Rev. Lett.} \textbf{112} 127203.

\bibitem{Starykh_Review} Starykh O A 2015 {\it Rep. Prog. Phys.} \textbf{78} 052502.

\bibitem{Sellmann} Sellmann D, Zhang X F and Eggert S 2015 {\it Phys. Rev. B} \textbf{91} 081104(R).

\bibitem{Coletta} Coletta T, T\'{o}th T A, Penc K and Mila F 2016 {\it Phys. Rev. B} \textbf{94} 075136.

\bibitem{Shirata} Shirata Y, Tanaka H, Matsuo A and Kindo K 2012 {\it Phys. Rev. Lett.} \textbf{108} 057205.

\bibitem{Susuki} Susuki T, Kurita N, Tanaka T, Nojiri H, Matsuo A, Kindo K and Tanaka H {\it Phys. Rev. Lett.} \textbf{110} 267201.

\bibitem{Okada} Okada K, Tanaka H, Kurita N, Yamamoto D, Matsuo A and Kindo K 2022 {\it Phys. Rev. B} \textbf{106} 104415.

\bibitem{Ono3} Ono T, Tanaka H, Aruga Katori H, Ishikawa F, Mitamura H and Goto T 2003 {\it Phys. Rev. B} \textbf{67} 104431.

\bibitem{Ono4} Ono T, Tanaka H, Kolomiyets O, Mitamura H, Goto T, Nakajima K, Oosawa A, Koike Y, Kakurai K, Klenke J, Smeibidle P and Mei{\ss}ner M 2004 {\it J. Phys.: Condens. Matter} \textbf{16} S773. 

\bibitem{Fortune} Fortune N A, Hannahs S T, Yoshida Y, Sherline T E, Ono T, Tanaka H and Takano Y 2009 {\it Phys. Rev. Lett.} \textbf{102} 257201.

\bibitem{Sera} Sera A, Kousaka Y, Akimitsu J, Sera M and Inoue K 2017 {\it Phys. Rev. B} \textbf{96} 014419.

\bibitem{Kojima} Kojima Y, Watanabe M., Kurita N, Tanaka H, Matsuo A, Kindo K and Avdeev M 2018 {\it Phys. Rev. B} \textbf{98} 174406.

\bibitem{Xing} Xing J, Sanjeewa L D, Kim J, Stewart G R, Podlesnyak A and Sefat A S 2019 {\it Phys. Rev. B} \textbf{100} 220407(R).

\bibitem{Ranjith} Ranjith K M, Luther S, Reimann T, Schmidt B, Schlender Ph, Sichelschmidt J, Yasuoka H, Strydom A M, Skourski Y, Wosnitza J, K{\"u}hne H, Doert Th and Baenitz M 2019 {\it Phys. Rev. B} \textbf{100} 224417.

\bibitem{Ding} Ding L, Manuel P, Bachus S, Gru{\ss}ler F, Gegenwart P, Singleton J, Johnson R D, Walker H C, Adroja D T, Hillier A D and Tsirlin A A 2019 {\it Phys. Rev. B} \textbf{100} 144432.

\bibitem{Yamamoto4} Yamamoto D, Sakurai T, Okuto R, Okubo S, Ohta H, Tanaka H and Uwatoko Y 2021 {\it Nat. Commun.} \textbf{12} 4263.

\bibitem{Zheng} Zheng W, Fj{\ae}restad J O, Singh R R P, McKenzie R H and Coldea R 2006 {\it Phys. Rev. B} \textbf{74} 224420.

\bibitem{Mezio} Mezio A, Sposetti C N, Manuel L O and Trumper A E 2011 {\it Europhys. Lett.} \textbf{94} 47001.

\bibitem{Ferrari} Ferrari F and Becca F 2019 {\it Phys. Rev. X} \textbf{9} 031026.

\bibitem{Zhang} Zhang C and Li T 2020 {\it Phys. Rev. B} \textbf{102} 075108.

\bibitem{Ghioldi} Ghioldi E A, Mezio A, Manuel L O, Singh R R P, Oitmaa J and Trumper A E 2015 {\it Phys. Rev. B} \textbf{91} 134423.

\bibitem{Ghioldi2} Ghioldi E A, Gonzalez M G, Zhang S-S, Kamiya Y, Manuel L O, Trumper A E and Batista C D 2018 {\it Phys. Rev. B} \textbf{98} 184403.

\bibitem{Ghioldi3} Ghioldi E A, Zhang S-S, Kamiya Y, Manuel L O, Trumper A E and Batista C D 2022 {\it Phys. Rev. B} \textbf{106} 064418.

\bibitem{Chi} Chi R, Liu Y, Wan Y, Liao H-J and Xiang T 2022 {\it Phys. Rev. Lett.} \textbf{129} 227201. 

\bibitem{Syromyatnikov2} Syromyatnikov A V 2022 {\it Phys. Rev. B} \textbf{105} 144414.

\bibitem{Starykh} Starykh O A, Chubukov A V and Abanov A G 2006 {\it Phys. Rev. B} \textbf{74} 180403(R).

\bibitem{Chernyshev} Chernyshev A L and Zhitomirsky M E 2009 {\it Phys. Rev. B} \textbf{79} 144416.

\bibitem{Mourigal} Mourigal M, Fuhrman W T, Chernyshev A L and Zhitomirsky M E 2013 {\it Phys. Rev. B} \textbf{88} 094407.

\bibitem{Verresen} Verresen R, Moessner R and Pollmann F 2019 {\it Nat. Phys.} \textbf{15} 750.

\bibitem{Ma} Ma J, Kamiya Y, Hong T, Cao H B, Ehlers G, Tian W, Batista C D, Dun Z L,  Zhou H D and Matsuda M 2016 {\it Phys. Rev. Lett.} \textbf{116} 087201.

\bibitem{Ito} Ito S, Kurita N, Tanaka H, Ohira-Kawamura S, Nakajima K, Itoh S, Kuwahara K and Kakurai K 2017 {\it Nat. Commun.} \textbf{8} 235.

\bibitem{Kamiya} Kamiya Y, Ge L, Hong T, Qiu Y, Quintero-Castro D L, Lu Z, Cao H B, Matsuda M, Choi E S, Batista C D, Mourigal M, Zhou H D and Ma J 2018 {\it Nat. Commun.} \textbf{9} 2666.

\bibitem{Macdougal} Macdougal D, Williams S, Prabhakaran D, Bewley R I, Voneshen D J and Coldea R 2020 {\it Phys. Rev. B} \textbf{102} 064421.

\bibitem{Zhou} Zhou H D, Xu C, Hallas A M, Silverstein H J, Wiebe C R, Umegaki I, Yan J Q, Murphy T P, Park J-H, Qiu Y, Copley J R D, Gardner J S and Takano Y 2012 {\it Phys. Rev. Lett.} \textbf{109} 267206.




\bibitem{Zeng1} Zeng C and Elser V 1990 {\it Phys. Rev. B} \textbf{42} 8436.

\bibitem{Sachdev} Sachdev S 1992 {\it Phys. Rev. B} \textbf{45} 12377.

\bibitem{Chalker} Chalker J T and Eastmond J F G 1992 {\it Phys. Rev. B} \textbf{46} 14201.

\bibitem{Elstner} Elstner N and Young A P 1994 {\it Phys. Rev. B} \textbf{50} 6871.

\bibitem{Zeng2} Zeng C and Elser V 1995 {\it Phys. Rev. B} \textbf{51} 8318.

\bibitem{Nakamura} Nakamura T and Miyashita S 1995 {\it Phys. Rev. B} \textbf{52} 9174.

\bibitem{Lecheminant} Lecheminant P, Bernu B, Lhuillier C, Pierre L and Sindzingre P 1997 {\it Phys. Rev. B} \textbf{56} 2521.

\bibitem{Waldtmann} Waldtmann Ch, Everts H-U, Bernu B, Lhuillier C, Sindzingre P, Lecheminant P and Pierre L 1998 {\it Eur. Phys. J. B} \textbf{2} 501.

\bibitem{Mila} Mila F 1998 {\it Phys. Rev. Lett.} \textbf{81} 2356.

\bibitem{Mambrini} Mambrini M and Mila F 2000 {\it Eur. Phys. J. B} \textbf{17} 651.

\bibitem{Syromyatnikov} Syromyatnikov A V and Maleyev S V 2002 {\it Phys. Rev. B} \textbf{66} 132408.

\bibitem{Jiang} Jiang H C, Weng Z Y and Sheng D N 2008 {\it Phys. Rev. Lett.} \textbf{101} 117203.

\bibitem{Nikolic} 
Nikolic P and Senthil T 2003 {\it Phys. Rev. B} \textbf{68} 214415.

\bibitem{Budnik} 
Budnik R and Auerbach A 2004 {\it Phys. Rev. Lett.} \textbf{93} 187205.

\bibitem{Singh1}
Singh R R P and Huse D A 2007 {\it Phys. Rev. B} \textbf{76} 180407(R).

\bibitem{Singh2}
Singh R R P and Huse D A 2008 {\it Phys. Rev. B} \textbf{77} 144415.

\bibitem{Yang}
Yang B-J, Kim Y B, Yu J and Park K 2008 {\it Phys. Rev. B} \textbf{77} 224424.

\bibitem{Hwang1} 
Hwang K, Kim Y B, Yu J and Park K 2011 {\it Phys. Rev. B} \textbf{84} 205133.

\bibitem{Hastings}
Hastings M. B. 2000 {\it Phys. Rev. B} \textbf{63} 014413.

\bibitem{Ran} Ran Y, Hermele M, Lee P A and Wen X-G 2007 {\it Phys. Rev. Lett.} \textbf{98} 117205.

\bibitem{Hermele} Hermele M, Ran Y, Lee P A and Wen X-G 2008 {\it Phys. Rev. B} \textbf{77} 224413.

\bibitem{Iqbal} Iqbal Y, Becca F, Sorella S and Poilblanc D 2013 {\it Phys. Rev. B} \textbf{87} 060405(R).

\bibitem{Liao} Liao H J, Xie Z Y, Chen J, Liu Z Y, Xie H D, Huang R Z, Normand B and Xiang T 2017 {\it Phys. Rev. Lett.} \textbf{118} 137202.
 
\bibitem{He} He Y C, Zaletel M P, Oshikawa M and Pollmann F 2017 {\it Phys. Rev. X} \textbf{7} 031020.

\bibitem{Zhu} Zhu W, Gong S S and Sheng D N 2019 {\it Proc. Natl. Acad. Sci. USA} \textbf{116} 5437.

\bibitem{Jiang2} Jiang S, Kim P, Han J H and Ran Y 2019 {\it SciPost Phys.} \textbf{7} 006.

\bibitem{Jiang3} Jiang H C, Weng Z Y and Sheng D N 2008 {\it Phys. Rev. Lett.} \textbf{101} 117203.

\bibitem{Yan} Yan S, Huse D A and White S R 2011 {\it Science} \textbf{332} 1173.

\bibitem{Depenbrock} Depenbrock S, McCulloch I P and Schollw{\"o}ck U 2012 {\it Phys. Rev. Lett.} \textbf{109} 067201.

\bibitem{Jiang4} Jiang H C, Wang Z and Balents L 2012 {\it Nat. Phys.} \textbf{8} 902.

\bibitem{Mei} Mei J W, Chen J Y, He H and Wen X-G 2017 {\it Phys. Rev. B} \textbf{95} 235107.


\bibitem{Chubukov2} ChubukovA 1992 {\it Phys. Rev. Lett.} \textbf{69} 832.

\bibitem{Reimers} Reimers J N and Berlinsky A J 1993 {\it Phys. Rev. B} \textbf{48} 9539.


\bibitem{Hida} Hida K 2001 {\it J. Phys. Soc. Jpn.} \textbf{70} 3673.

\bibitem{Honecker_kagome} Honecker A, Schulenburg J and Richter J 2004 {\it J. Phys.: Condens. Mater} \textbf{16} S749.

\bibitem{Nishimoto} Nishimoto S, Shibata N and Hotta C 2013 {\it Nat. Commun.} \textbf{4} 2287.

\bibitem{Capponi} Capponi S, Derzhko O, Honecker A, L{\"a}uchli A M and Richter J 2013 {\it Phys. Rev. B} \textbf{88} 144416.

\bibitem{Schnack} Schnack J, Schulenburg J and Richter J 2018 {\it Phys. Rev. B} \textbf{98} 094423.

\bibitem{Plat} Plat X, Momoi T and Hotta C 2018 {\it Phys. Rev. B} \textbf{98} 014415.

\bibitem{Morita_kagome} K. Morita, Phys. Rev. B \textbf{108}, 184405 (2023). 

\bibitem{Yoshida} Yoshida H 2022 {\it J. Phys. Soc. Jpn.} \textbf{91} 101003.

\bibitem{Mueller} 
M\"{u}ller M and M\"{u}ller B G 1995 {\it Z. Anorg. Allg. Chem.} \textbf{621} 993.

\bibitem{Hiroi} 
Hiroi Z, Hanawa M, Kobayashi N, Nohara M, Takagi H, Kato Y and Takigawa M 2001 {\it J. Phys. Soc. Jpn.} \textbf{70} 3377.

\bibitem{Shores} 
Shores M P, Nytko E A, Bartlett B M and Nocera D G 2005 {\it J. Am. Chem. Soc.} \textbf{127} 13462. 

\bibitem{Okamoto} 
Okamoto Y, Yoshida H and Hiroi Z 2009 {\it J. Phys. Soc. Jpn.} \textbf{78} 033701.

\bibitem{Fak}
F{\aa}k B, Kermarrec E, Messio L, Bernu B, Lhuillier C, Bert F, Mendels P, Koteswararao B, Bouquet F, Ollivier J, Hillier A D, Amato A, Colman R H and Wills A S 2012 {\it Phys. Rev. Lett.} \textbf{109} 037208.

\bibitem{Ishikawa}
Ishikawa H, Okamoto Y and Hiroi Z 2013 {\it J. Phys. Soc. Jpn.} \textbf{82} 063710.

\bibitem{Fujihala}
Fujihala M, Zheng W-G, Morodomi H, Kawae T, Matsuo A, Kindo K and Watanabe I 2014 {\it Phys. Rev. B} \textbf{89} 100401(R).

\bibitem{Sun}
Sun W, Huang Y-X, Nokhrin S, Pan Y and Mi J-X 2016 {\it J. Mater. Chem. C} \textbf{4} 8772.

\bibitem{Sun2}
Sun W, Huang Y-X, Pan Y and Mi J-X 2016 {\it Phys. Chem. Minerals} \textbf{43} 127.

\bibitem{Yoshida2}
Yoshida H, Noguchi N, Matsushita Y, Ishii Y, Ihara Y, Oda M, Okabe H, Yamashita S, Nakazawa Y, Takata A, Kida T, Narumi Y and Hagiwara M 2017 {\it J. Phys. Soc. Jpn.} \textbf{86} 033704.

\bibitem{Puphal}
Puphal P, Bolte M, Sheptyakov D, Pustogow A, Kliemt K, Dressel M, Baenitz M and Krellner C 2017 {\it J. Mater. Chem. C} \textbf{5} 2629.
\bibitem{Feng} Feng Z, Li Z, Meng X, Yi W, Wei Y, Zhang J, Wang Y-C, Jiang W, Liu Z, Li S, Liu F, Luo J, Li S, Zheng G-Q, Meng Z Y, Mei J-W and Shi Y 2017 {\it Chin. Phys. Lett.} \textbf{34} 077502.
\bibitem{Wang} Wang R, Li X, Han X, Lin J, Wang Y, Qian T, Ding H, Shi Y and Liu X 2021 {\it Chin. Phys. B} \textbf{30} 046102.

\bibitem{Mendels} 
Mendels P, Bert F, de Vries M A, Olariu A, Harrison A, Duc F, Trombe J C, Lord J S, Amato A and Baines C 2007 {\it Phys. Rev. Lett.} \textbf{98} 077204.

\bibitem{Helton}
Helton J S, Matan K, Shores M P, Nytko E A, Bartlett B M, Yoshida Y, Takano Y, Suslov A, Qiu Y, Chung J-H, Nocera D G and Lee Y S 2007 {\it Phys. Rev. Lett.} \textbf{98} 107204.

\bibitem{Lee} 
Lee S-H, Kikuchi H, Qiu Y, Lake B, Huang Q, Habicht K and Kiefer K 2007 {\it Nat. Mater.} \textbf{6} 853.

\bibitem{Bert} 
Bert F, Nakamae S, Ladieu F, L'H\^{o}te D, Bonville P, Duc F, Trombe J-C and Mendels P 2007 {\it Phys. Rev. B} \textbf{76} 132411.

\bibitem{Imai}
Imai T, Nytko E A, Bartlett B M, Shores M P and Nocera D G 2008 {\it Phys. Rev. Lett.} \textbf{100} 077203.

\bibitem{Olariu}
Olariu A, Mendels P, Bert F, Duc F, Trombe J C, de Vries M A and Harrison A 2008 {\it Phys. Rev. Lett.} \textbf{100} 087202.

\bibitem{Vries}
de Vries M A, Kamenev K V, Kockelmann W A, Sanchez-Benitez J and Harrison A 2008 {\it Phys. Rev. Lett.} \textbf{100} 157205.

\bibitem{Mendels2} Mendels P and Bert F. 2010 {\it J. Phys. Soc. Jpn.} \textbf{79} 011001.

\bibitem{Han2} Han T H, Helton J S, Chu S, Prodi A, Singh D K, Mazzoli C, M\"{u}ller P, Nocera D G and Lee Y S 2011 {\it Phys. Rev. B} \textbf{83} 100402(R).

\bibitem{Fu} Fu M X, Imai T, Han T H and Lee Y S 2015 {\it Science} \textbf{350} 655.

\bibitem{Zorko} Zorko A, Herak M, Gomil\v{s}ek M, van Tol J, Vel\'azquez M, Khuntia P, Bert F and Mendels P 2017 {\it Phys. Rev. Lett.} \textbf{118} 017202.

\bibitem{Khuntia} Khuntia P, Velazquez M, Barth\'elemy Q, Bert F, Kermarrec E, Legros A, Bernu B, Messio L, Zorko A and Mendels P 2020 {\it Nat. Phys.} \textbf{16} 469 .

\bibitem{Han} Han T H, Helton J S, Chu S, Nocera D G, Rodriguez-Rivera J A, Broholm C and Lee Y S 2012 {\it Nature} \textbf{492} 406.

\bibitem{Han3} Han T H, Norman M R, Wen J-J, Rodriguez-Rivera J A, Helton J S, Broholm C and Lee Y S 2016 {\it Phys. Rev. B} \textbf{94} 060409(R).

\bibitem{Punk} Punk M, Chowdhury D and Sachdev S 2014 {\it Nat. Phys.} \textbf{10} 289.

\bibitem{Zhang2} Zhang C and Li T 2020 {\it Phys. Rev. B} \textbf{102} 195106. 

\bibitem{Ferrari2} Ferrari F, Parola A and Becca F 2021 {\it Phys. Rev. B} \textbf{103} 195140.

\bibitem{Prelovsek} Prelov\v{s}ek P, Gomil\v{s}ek M, Arh T and Zorko A. 2021 {\it Phys. Rev. B} \textbf{103} 014431.

\bibitem{Watanabe} Watanabe K, Kawamura H, Nakano H and Sakai T {\it J. Phys. Soc. Jpn.} \textbf{83} 034714.

\bibitem{Kawamura2014} Kawamura H, Watanabe K and Shimokawa T {\it J. Phys. Soc. Jpn.} \textbf{83} 103704.

\bibitem{Kawamura2019} Kawamura H and Uematsu K {\it J. Phys.: Condens. Matter} \textbf{31} 504003.

\bibitem{Shimokawa} Shimokawa T, Watanabe K and Kawamura H 2015 {\it Phys. Rev. B} \textbf{92} 134407.

\bibitem{Boldrin} Boldrin D, F{\aa}k B, Can\'{e}vet E, Ollivier J, Walker H C, Manuel P,  Khalyavin D D and Wills A S 2018 {\it Phys. Rev. Lett.} \textbf{121} 107203.


\bibitem{Ono} Ono T, Morita K, Yano M, Tanaka H, Fujii K, Uekusa H, Narumi Y and Kindo K. 2009 {\it Phys. Rev. B} \textbf{79} 174407.

\bibitem{Ono2} Ono T, Matan K, Nambu Y, Sato T J, Katayama K, Hirata S and Tanaka H 2014 {\it J. Phys. Soc. Jpn.} \textbf{83} 043701.

\bibitem{Matan3} Matan K, Ono T, Gitgeatpong G, de Roos K, Miao P, Torii S, Kamiyama T, Miyata A, Matsuo A, Kindo K, Takeyama S, Nambu Y, Piyawongwatthana P, Sato T J and Tanaka H 2019 {\it Phys. Rev. B} \textbf{99} 224404.

\bibitem{Saito}
Saito M, Takagishi R, Kurita N, Watanabe M, Tanaka H, Nomura R, Fukumoto Y, Ikeuchi K and Kajimoto R 2022 {\it Phys. Rev. B} \textbf{105} 064424.

\bibitem{Matan4}
Matan K, Ono T, Ohira-Kawamura S, Nakajima K, Nambu Y and Sato T J 2022 {\it Phys. Rev. B} \textbf{105} 134403.

\bibitem{Ferrari3} 
Ferrari F, Niu S, Hasik J, Iqbal Y, Poilblanc D and Becca F 2023 {\it SciPost Phys.} \textbf{14} 139.

\bibitem{Kogure}
Kogure S, Takeda M, Morita K, Fukumoto Y, Saito M and Tanaka H 2023 {\it J. Phys. Soc. Jpn.} \textbf{92} 1137.

\bibitem{Morita} Morita K, Yano M, Ono T, Tanaka H, Fujii K, Uekusa H, Narumi Y and Kindo K 2008 {\it J. Phys. Soc. Jpn.} \textbf{77} 043707.

\bibitem{Matan} Matan K, Ono T, Fukumoto Y, Sato T J, Yamaura J, Yano M, Morita K and Tanaka H 2010 {\it Nat. Phys.} \textbf{6} 865.

\bibitem{Matan2} Matan K, Nambu Y, Zhao Y, Sato T J, Fukumoto Y, Ono T, Tanaka H, Broholm C, Podlesnyak A and Ehlers G 2014 {\it Phys. Rev. B} \textbf{89} 024414.

\bibitem{Elhajal} 
Elhajal M, Canals B and Lacroix C 2002 {\it Phys. Rev. B} \textbf{66} 014422.

\bibitem{Downie2} Downie L J, Black C, Ardashnikova E I, Tang C C, Vasiliev A N, Golovanov A N, Berdonosov P S, Dolgikh V A and Lightfoot P 2014 {\it CrystEngComm} \textbf{16} 7419.

\bibitem{Satija} Satija S K, Cox J D, Shirane G, Yoshizawa H and Hirakawa K 1980 {\it Phys. Rev. B} \textbf{21} 2001.

\bibitem{Nagler} Nagler S E, Tennant D A, Cowley R A, Perring T G and Satija S K 1991 {\it Phys. Rev. B} \textbf{44} 12361.

\bibitem{Amemiya1} Amemiya T, Yano M, Morita K, Umegaki I, Ono T, Tanaka H, Fujii K and Uekusa H 2009 {\it Phys. Rev. B} \textbf{80} 100406(R).

\bibitem{Umegaki1} Umegaki I, Tanaka H, Ono T, Uekusa H and Nojiri H 2009 {\it Phys. Rev. B} \textbf{79} 184401. 

\bibitem{Umegaki2} Umegaki I, Tanaka H, Kurita N, Ono T, Laver M, Niedermayer Ch, R\"{u}egg Ch, Ohira-Kawamura S, Nakajima K and Kakurai K 2015 {\it Phys. Rev. B} \textbf{92} 174412.

\bibitem{Goodenough} Goodenough J B 1958 {\it J. Phys. Chem. Solids} \textbf{6} 287.

\bibitem{Kanamori} Kanamori J 1959 {\it J. Phys. Chem. Solids} \textbf{10} 87.

\bibitem{Downie} Downie L J, Thompson S P, Tang C C, Parsons S and Lightfoot P 2013 {\it CrystEngComm} \textbf{15} 7426.

\bibitem{Rigol} Rigol M and Singh R R P 2007 {\it Phys. Rev. B} \textbf{76} 184403.

\bibitem{Misguich2} 
Misguich G and Sindzingre P 2007 {\it Eur. Phys. J. B} \textbf{59} 305.

\bibitem{Katayama} Katayama K, Kurita N and Tanaka H 2015 {\it Phys. Rev. B} \textbf{91} 214429.
 
\bibitem{Cepas} 
C\'{e}pas O, Fong C M, Leung P W and Lhuillier C 2008 {\it Phys. Rev. B} \textbf{78} 140405(R).

\bibitem{Rousochatzakis} Rousochatzakis I, Manmana S R, L\"{a}uchli A M, Normand B and Mila F 2009 {\it Phys. Rev. B} \textbf{79} 214415.

\bibitem{Hering} Hering M and Reuther J 2017 {\it Phys. Rev. B} \textbf{95} 054418.

\bibitem{Lee2} Lee C-Y, Normand B and Kao Y-J 2018 {\it Phys. Rev. B} \textbf{98} 224414.

\bibitem{Zorko2} Zorko A, Nellutla S, van Tol J, Brunel L C, Bert F, Duc F, Trombe J-C, de Vries M A, Harrison A and Mendels P 2008 {\it Phys. Rev. Lett.} \textbf{101} 026405.

\bibitem{Arh} Arh T, Gomil\v{s}ek M, Prelov\v{s}ek P, Pregelj M, Klanj\v{s}ek M, Ozarowski A, Clark S J, Lancaster T, Sun W, Mi J-X and Zorko A 2020 {\it Phys. Rev. Lett.} \textbf{125} 027203.
\bibitem{Grbic} Grbi\'{c} M S, Kr\"{a}mer S, Berthier C, Trousselet F, C\'{e}pas O, Tanaka H and Horvati\'{c} M 2013 {\it Phys. Rev. Lett.} \textbf{110} 247203.

\bibitem{Suzuki} Suzuki T, Katayama K, Watanabe I and Tanaka H 2020 {\it J. Phys. Soc. Jpn.} \textbf{89} 074701.

\bibitem{Hayashida} Hayashida S, Matsumoto M, Hagihala M, Kurita N, Tanaka H, Itoh S, Hong T, Soda M, Uwatoko Y and Masuda T 2019 {\it Science Adv.} \textbf{5} eaaw5639.

\bibitem{Matan5} Matan K, Grohol D, Nocera D G, Yildirim T, Harris A B, Lee S H, Nagler S E and Lee Y S 2006 {\it Phys. Rev. Lett.} \textbf{96} 247201.

\bibitem{Yildirim} Yildirim T and Harris A B 2006 {\it Phys. Rev. B} \textbf{73} 214446. 

\bibitem{Lake} Lake B, Tennant D A, Caux J-S, Barthel T, Schollw\"{o}ck U, Nagler S E and Frost C D 2013 {\it Phys. Rev. Lett.} \textbf{111} 137205.

\bibitem{dCP} des Cloizeaux J and Pearson J J 1962 {\it Phys. Rev.} \textbf{128} 2131.

\bibitem{Faddeev} Faddeev L D and Takhtajan L A 1981 {\it Phys. Lett.} \textbf{85A} 375.

\bibitem{Muller2} M\"{u}ller G, Thomas H, Beck H and Bonner J C 1981 {\it Phys.
Rev. B} \textbf{24} 1429.


\bibitem{Yang2}
Yang B-J and Kim Y B 2009 {\it Phys. Rev. B} \textbf{79} 224417.

\bibitem{Hwang2} Hwang K, Park K and Kim Y B 2012 {\it Phys. Rev. B} \textbf{86} 214407.

\bibitem{Khatami} Khatami E, Singh R R P and Rigol M. 2011 {\it Phys. Rev. B} \textbf{84} 224411.


\end{thebibliography}
\end{document}